\documentclass[12pt, a4paper]{article}
\usepackage{arxiv}
\usepackage{tikz}
\usetikzlibrary{arrows.meta, positioning, shapes.geometric}
\usepackage[utf8]{inputenc} 
\usepackage[T1]{fontenc}   
\usepackage[hidelinks]{hyperref}       
\usepackage{url}            
\usepackage{booktabs}     
\usepackage{amsfonts}      
\usepackage{nicefrac}       
\usepackage{microtype}     
\usepackage{lipsum}
\usepackage{graphicx}
\usepackage{bbm}
\usepackage{amsmath,amssymb}
\graphicspath{ {./images/} }
\usepackage{float}
\usepackage{natbib}
\usepackage{multirow}
\usepackage[linesnumbered,algoruled,boxed,lined]{algorithm2e}
\usepackage{tcolorbox}
\usepackage{enumitem}
\usepackage{diagbox} 
\usepackage{array}  
\usepackage{ragged2e}
\usepackage{caption}
\usepackage{orcidlink} 
\usepackage{booktabs}
\usepackage{setspace}

\newtheorem{result}{Result}

\usepackage[linesnumbered,algoruled,boxed,lined]{algorithm2e}

\title{Inference from multivariate differential
recruitment in respondent-driven sampling data}

\author{
 Vanesa Reinoso \orcidlink{0009-0008-3803-2880}\\
  Department of Statistics\\
  Pontificia Universidad Cat{\'o}lica de Chile\\
  Santiago, Chile \\
  \texttt{vcreinoso@uc.cl} \\
  \And
 Danilo Alvares \orcidlink{0000-0003-3764-0397}\\
  MRC Biostatistics Unit\\
  University of Cambridge\\
  Cambridge, UK \\
  \texttt{danilo.alvares@mrc-bsu.cam.ac.uk} \\
  \And
 Jonathan Acosta \orcidlink{0000-0001-6323-9746} \\
  Department of Statistics\\
  Pontificia Universidad Cat{\'o}lica de Chile\\
  Santiago, Chile \\
  \texttt{jonathan.acosta@uc.cl} \\
  \And
 Isabelle S. Beaudry \orcidlink{0000-0001-7550-5570} \\
  Department of Mathematics and Statistics\\
  Mount Holyoke College\\
  Massachusetts, USA \\
  \texttt{ibeaudry@mtholyoke.edu} \\
}

\begin{document}
\maketitle

\begin{abstract}
Respondent-Driven Sampling (RDS) is a chain-referral design used for collecting data from hidden or hard-to-reach populations through their social networks. In RDS, respondents recruit their peers from the population of interest. As such, inference with RDS data commonly relies on estimated sampling probabilities derived from specific recruitment assumptions. Early literature assumes random recruitment, which is often unrealistic because individuals may recruit based on their personal preferences. This behavior is known as Differential Recruitment (DR). Recent works have incorporated univariate categorical DR in the estimation procedures. The main objective of this paper is to introduce Multivariate Differential Recruitment (MDR), a framework that incorporates multiple simultaneous covariates, both categorical and continuous, into the sampling representation. We model RDS as a Markov process with transition probabilities that depend on continuous or categorical variables associated with nodes or their ties. We then extend various prevalence estimators to this multivariate framework and implement a slightly modified neighborhood bootstrap for variance estimation. The proposed methodology is assessed through simulation studies for a range of network and sampling features. It is applied to an RDS study conducted among the adult Venezuelan population living in the Metropolitan Region of Santiago, Chile.
\end{abstract}

\section{Introduction}

Respondent-Driven Sampling (RDS) is a chain-referral sampling design used to access hard-to-reach or hidden populations for whom conventional sampling frames are unavailable \citep{Heckathorn1997,Heckathorn2002RDS_Extensions, Johnston2010, Gile2018}. Public health organizations were among the early adopters of RDS, particularly to sample populations most at risk for HIV/AIDS \citep{Magnani2005, Lansky2007, Arayasirikul2015}. Over time, the use of RDS has expanded to a broader range of populations and research objectives. For example, it has been used to reach women and girls in refugee settings who have experienced abortion to estimate abortion incidence \citep{Jayaweera2025}, people experiencing homelessness to study illicit substance use \citep{Assaf2025}, and immigrant populations to assess political participation \citep{Takahashi2025, Leal2025}. Although these applications illustrate the versatility of RDS, the validity of statistical inference based on such data highly depends on the accuracy of the recruitment assumptions. In this paper, we introduce Multivariate Differential Recruitment (MDR), a novel framework that accounts for multiple, simultaneous sources of recruitment bias.

Under Respondent-Driven Sampling (RDS), participants are tasked with recruiting a limited number of peers from their personal networks within the target population. Consequently, aside from the initial seeds selected by the investigators, the composition of the final sample is determined by participants' recruitment behaviors \citep{Beaudry2020, Rudolph2024}. Early RDS literature typically assumes that this recruitment process occurs at random. Specifically, participants are presumed to recruit alters in their network with equal probability \citep{Heckathorn2002, Salganik2004, Gile2018}.

An increasing body of literature has assessed the plausibility of the random peer-recruitment assumption \citep{Goel2009, Tourangeau2014, Gile2015, Barash2016, Gile2018, Li2018, Wang2024}. Furthermore, several studies have examined the consequences of relying on this unrealistic assumption for statistical inference, showing that estimates of population proportions can be substantially biased when it is violated \citep{Johnston2008, Gile2010, McCreesh2012, Liu2012, Yamanis2013, Wirtz2021, Avery2023, Rudolph2024}.

Before the development of alternative sampling formulations that explicitly account for non-random recruitment patterns, some scholars proposed estimators that are less sensitive to departures from the random recruitment assumption \citep{Lu2013, Roch2018, Fellows2019, Fellows2022}.

More recently, some work has introduced alternative representations of the RDS that incorporate unequal recruitment probabilities and explicitly characterize recruitment behavior within the sampling model \citep{Shi2019, Beaudry2020}. These approaches acknowledge that participants may prefer recruiting certain types of peers and incorporate these preferences directly into the inferential framework to mitigate the associated bias. This type of non-random recruitment behavior is commonly referred to as Differential Recruitment (DR) \citep{Heckathorn2007, Heckathorn2011, Tomas2011, Shi2019, Beaudry2020}. DR may occur when respondents systematically recruit peers based on a characteristic of the potential recruit (e.g.,~has completed high school), a shared attribute between the recruiter and the potential recruit (e.g.,~both are currently employed), or a feature of their relationship (e.g.,~relatives versus acquaintance). Although incorporating DR into estimation procedures has been shown to improve prevalence estimates \citep{Shi2019, Beaudry2020}, existing approaches are limited in that they model recruitment patterns with respect to only a single categorical variable at a time.

This paper's main contribution is to address this limitation by introducing MDR. This method improves statistical inference by modeling non-random recruitment as a function of multiple covariates, whether categorical or continuous. For example, the MDR sampling model for RDS may simultaneously include alters' occupations, the age difference between the participant and their alters, and the frequency of communication between them. If these factors influence participation in the RDS study, including them in the sampling model and estimation procedure may help reduce bias.

To account for MDR in the inference, we extend the work of \cite{Beaudry2020} to a multivariate framework. Specifically, we model RDS as a Markov chain (MC) whose transition probabilities depend on covariates, with parameters that weight their influence. Under some conditions, we establish that this MC admits a unique stationary distribution, from which individual sampling probabilities are derived. These probabilities are then used to update the prevalence estimators of \cite{Volz2008} and \cite{Lu2013}, which were previously extended to the DR setting by \cite{Beaudry2020} to incorporate MDR explicitly. Finally, we make a minor modification to the bootstrap methodology of \cite{Yauck2022} to estimate the variance of the resulting estimators.

We evaluate the performance of the proposed methods through simulation studies, demonstrating that in the presence of network homophily and MDR, the extended estimators outperform those based on the assumptions of random recruitment and univariate DR.

The remainder of the paper is organized as follows. Section~\ref{sec:2} introduces Respondent-Driven sampling, reviews existing estimators, and discusses their associated variance estimation methods. Section~\ref{sec:MDR} presents the proposed MDR model, along with extended prevalence estimators and their corresponding variance estimators. Section~\ref{sec:4} describes the simulation study conducted to assess and compare the performance of the estimators. Section~\ref{sec:5} illustrates the proposed methodology using a study carried out with Venezuelans residing in the Metropolitan Region of Santiago, Chile. Finally, Section~\ref{sec:6} concludes with a discussion of the results.

\section{Respondent-driven sampling} \label{sec:2}

Developing a conventional sampling frame is often impractical for hidden or hard-to-reach populations. Respondent-Driven sampling (RDS) \citep{Heckathorn1997} addresses this challenge by leveraging the social connections that link individuals within the target population. Rather than selecting participants directly from a list, the sample grows through peer referral, in which participants recruit eligible individuals from their own social network. As a result, the sample is shaped by both the underlying social structure and participants' recruitment behavior.

Recruitment proceeds in waves, beginning with a small number of initial participants, commonly referred to as {\it seeds}, selected through convenience sampling. After completing the survey, each seed receives a limited number of coupons to invite peers from the same population. These individuals then recruit others from their network, thereby extending the recruitment chains. The sample grows through successive waves until it reaches the pre-established sample size.

\subsection{Notation}

Following \cite{Beaudry2020}, consider a hidden population of size $N$, with individuals indexed by $i \in \{1, \dots, N\}$. The social structure is represented by a simple undirected network, where nodes correspond to individuals and ties represent social connections. Let $\boldsymbol{Y} \in \{0,1\}^{N \times N}$ denote the adjacency matrix of this network, where $y_{ij} = y_{ji} = 1$ if a tie exists between individuals $i$ and $j$, and $y_{ij} =y_{ji} = 0$ otherwise. The diagonal entries satisfy $y_{ii} = 0$ for all $i$, reflecting the absence of self-ties. 

Associated with each individual $i$ is a fixed binary response variable $z_i \in \{0,1\}$, which, for example, may indicate whether individual $i$ is infected with a disease ($z_i = 1$) or not ($z_i = 0$). We collect these response variables into a population vector $\boldsymbol{z} = (z_1, \dots, z_N)^\top$. For simplicity, we will use the terms response variable and infection status interchangeably throughout the paper. This notation facilitates the description of the main inferential goal in the following sections, that is, to estimate the prevalence of a disease:
\begin{equation}
\mu = N_1 / N,\label{eq:mu1}
\end{equation}
where $N_1 = \sum_{i=1}^{N} z_i$ is the number of infected individuals by such a disease in the population. Respondents must also report their degree in the RDS survey, defined as the total number of connections in their own network, $d_i = \sum_{j=1}^{N} y_{ij}$. Although degrees may be subject to measurement error in practice, for this work, we assume they are measured without error. Finally, we define an $N$-dimensional vector $\boldsymbol{S}$, where the  $i$-th entry equals 1 if individual $i$ is included in the sample, and 0 otherwise.

\subsection{Prevalence estimators} \label{sec:prevalence}

In this section, we review existing prevalence estimators. Unlike traditional sampling designs, RDS sampling probabilities are neither fixed in advance nor identifiable from the observed sample data alone. Consequently, the RDS literature focuses on modeling the sampling mechanism to estimate the pseudo-sampling probabilities $\hat{p}_i$. The estimators we review can be broadly classified into two categories based on their modeling choices: those that assume random recruitment and those that explicitly account for differential recruitment.

We begin with the random recruitment estimators, proposed by \cite{Salganik2004}, \cite{Volz2008}, and \cite{Lu2013}, and denoted by $\hat{\mu}_{SH}^{I}$, $\hat{\mu}_{VH}^{II}$, and $\hat{\mu}_{Lu}^{ego}$, respectively. Then, we present the differential recruitment estimators, introduced by \cite{Beaudry2020}, and denoted by $\hat{\mu}_{DR}^{I}, \hat{\mu}_{DR}^{II}$ and $\hat{\mu}_{DR}^{ego}$.

\subsubsection{Random recruitment prevalence estimators}

RDS is frequently modeled as a Markov chain (MC), where the state space consists of the network's nodes, and the chain's transitions represent the recruitment process \citep{Salganik2004}. Under the random recruitment scheme, the key assumption is that each participant selects a contact uniformly at random. Formally, the recruitment process is represented by a transition matrix $\boldsymbol{P}$, where the entry $P_{ij} = y_{ij}/d_i$ gives the probability that individual $i$ recruits individual $j$. Therefore, the probability of recruiting an alter depends on the recruiter's degree and is identical across all of $i$'s contacts. \cite{Salganik2004} proved that, under mild conditions, this MC has a unique stationary distribution, directly proportional to the individuals' degree, i.e., $\pi_{i} \propto d_{i}$ $\forall$ $i \in \{1, \ldots, N\}$. This MC stationary distribution is used to estimate the selection probabilities of the estimators presented in this section.

The first estimator we consider is the VH estimator ($\hat{\mu}_{VH}^{II}$), which was introduced by \cite{Volz2008}. It belongs to the class of Hájek-type estimators (HT), defined as the ratio of Hansen-Hurwitz estimators \citep{Hansen1943}, in which the true sampling probabilities $p_i$ are replaced by their corresponding estimates $\hat{p}_i$. The general form of the HT prevalence estimator is given by:
\begin{equation} \label{eq:ht}
\hat{\mu}_{HT} =
\frac{\sum_{i=1}^{N} S_i z_i / \hat{p}_i}
{\sum_{i=1}^{N} S_i / \hat{p}_i}.
\end{equation} 

Under the discussed MC, the sampling probabilities are assumed to equal the stationary distribution and, therefore, are proportional to the nodes' degrees. Although the sampling probabilities are identifiable only up to a constant of proportionality from the observed data, this constant cancels in the ratio. Thus, the VH estimator substitutes $\hat{p}_i$ by $d_i$ into Equation~\eqref{eq:ht} to obtain: 
\begin{equation} \label{eq:VH}
\hat{\mu}_{VH}^{II} = \frac{\sum_{i=1}^{N} S_i z_i / d_i}{\sum_{i=1}^{N} S_i/d_i}.
\end{equation}

The second estimator we discuss was introduced by \cite{Salganik2004} and is referred to as the SH estimator ($\hat{\mu}_{SH}^{I}$). The authors demonstrated that the true population prevalence for an undirected network can be expressed as:
\begin{equation}
\mu = \frac{C_{01} D_0}{C_{01} D_0 + C_{10} D_1},
\label{eq:mu}
\end{equation}
where $C_{kl}$ denotes the proportion of ties from group $k$ to group $l$ ($k,l \in \{0,1\}$), and $D_k$ is the average degree of individuals in group $k$. In the context of RDS, group~1 may represent infected individuals, and group~0, uninfected individuals. The intuition behind this result is that, in an undirected network, the number of cross-group connections from uninfected to infected individuals ($T_{01}$) must equal the number of connections from infected to uninfected individuals ($T_{10}$), such that:
\begin{equation}
T_{01} = \sum_{i,j} (1-z_i) z_j y_{ij} = \sum_{i,j} z_i (1-z_j) y_{ij} = T_{10}.
\end{equation}

This reciprocity establishes a relationship between the relative sizes of the two groups, $N_1/N_0$, and their average degrees and cross-group mixing proportions, $C_{01} D_0$ and $C_{10} D_1$, thereby allowing the prevalence to be expressed in terms of these network quantities.

The quantities $D_k$ and $C_{k,1-k}$ for $k \in \{0,1\}$ in Equation \eqref{eq:mu} are population-level parameters and must be estimated through observable variables. In particular, the SH estimator substitutes them with the following estimates:
\begin{equation}
\widehat{C}_{k,1-k}^{SH} 
= \frac{\sum_{i,j} S_{ij} \, \mathbbm{1}\left[z_i = k, \, z_j = 1-k \right]}
       {\sum_{i,j} S_{ij} \, \mathbbm{1}\left[ z_i = k \right]}, \quad \widehat{D}_{k} = \frac{\sum_{i=1}^{N}S_{i}\mathbbm{1}{[z_{i} = k]} }{{\sum_{i=1}^{N}S_{i}\mathbbm{1}{[z_{i} = k]}/d_{i}}} ~ k \in \{0,1\},
\label{eq:D}
\end{equation}
where $S_{ij} = 1$ if individual $i$ recruited individual $j$ in the RDS sample, and $0$ otherwise, and $\mathbbm{1}[\cdot]$ is the indicator function. Thus, $\widehat{C}_{k,1-k}^{SH}$ represents the sample proportion of cross-group recruitment from individuals with infection status $k$ to those with status $1-k$. Under the random recruitment assumption, it may be used as an estimate for $C_{k,1-k}$, the proportion of cross-group ties. As for $D_{k}$, it is estimated through a Hájek-type estimator.

Finally, the third estimator $\hat{\mu}_{Lu}^{ego}$, developed by \cite{Lu2013}, shares the same structural form as the SH estimator in Equation~\eqref{eq:mu}. However, rather than estimating $C_{k,1-k}$ directly from the observed recruitment, \citet{Lu2013} uses a Hájek-type estimator that incorporates information from the local network composition, such that:
\begin{equation}
\widehat{C}_{k,1-k}^{ego}=\left[\sum_{i=1}^{N}\frac{S_{i}d_{i(1-k)}^{z}\mathbbm{1}[z_{i} = k]}{d_{i}}\right]\left[\sum_{i=1}^{N}\mathbbm{1}[z_{i}=k]S_{i}\right]^{-1},  
\end{equation}
where $d_{ik}^{z} = \sum_{j=1}^{N} y_{ij} \mathbbm{1}[z_{j} = k]$ denotes the number of connections from node $i$ to individuals with infection status $k$. 

Although both the Lu and SH estimators share the same overall structure and use the same Hájek-type estimator for $\widehat{D}_k$, the methodological difference in estimating $C_{01}$ and $C_{10}$ has been shown to substantially improve the accuracy of the prevalence point estimate, even in the presence of differential recruitment \citep{Lu2013, Verdery2015, Beaudry2020}.

\subsubsection{Differential recruitment prevalence estimators} \label{sec:DRprev}

In this section, we discuss the prevalence estimators developed by \cite{Beaudry2020} to consider differential recruitment (DR) on a binary variable. DR arises when individuals do not recruit randomly but instead select peers with unequal probabilities. As noted by the authors, DR can manifest in several ways. In this paper, we focus exclusively on {\it between-group} DR, which occurs when participants preferentially recruit peers who exhibit a specific feature.

To quantify the magnitude or intensity of DR favoring the recruitment of individuals with a given attribute, the authors introduced the parameter $\phi > 0$, defined as the ratio of the probability of recruiting an individual with the DR characteristic ($u = 1$) to that of recruiting one without it ($u = 0$). For example, suppose the DR characteristic indicates whether an individual visits a specific outreach clinic, where $u = 1$ if they use the clinic and $u = 0$ otherwise. If a recruiter is connected to two peers, $j_1$ and $j_2$, where $u_{j_1} = 1$ and $u_{j_2} = 0$, the recruiter is $\phi$ times more likely to select $j_1$ than $j_2$. A value of $\phi = 1$ indicates that the recruiter is equally likely to select either peer, representing a random recruitment regime.

The parameter $\phi$ is used to specify the transition probabilities of the Markov process such that:
\begin{equation}
    P_{ij}^{\text{DR}} = \frac{y_{ij} \phi^{u_j}}{\sum_{l=1}^{N} y_{il} \phi^{u_l}}, \quad i, j = 1, \ldots, N.
\label{eq:drprobs}
\end{equation}

Under certain conditions, this MC admits a unique stationary distribution:
\begin{equation}
    \pi_i^{\text{DR}} \propto \phi^{u_i} \big(\phi d_{i1}^{u} + d_{i0}^{u}\big), \quad i = 1,\ldots,N,
\label{eq:drstat}
\end{equation}
where $d_{ik}^{u} = \sum_{j=1}^{N} y_{ij} \mathbbm{1}[u_j = k]$ denotes the number of connections from individual $i$ to nodes with $u = k$, for $k \in \{0,1\}$.  Intuitively, the long-run probability that the chain visits node $i$ is higher if that node possesses the DR characteristic ($\phi^{u_i} $) and is connected to many nodes with this characteristic ($\phi d_{i1}^{u}$).

As in the random recruitment regime, the stationary distribution may be used to approximate RDS sampling weights. However, since $\phi$ is unknown, it must first be estimated. The authors propose estimating this parameter by maximizing the following likelihood function:
\begin{equation}
\label{eq:likdr}
\mathcal{L}(\phi \mid \boldsymbol{R}, \boldsymbol{D}_{1}^{u}, \boldsymbol{D}_{0}^{u})
= \prod_{j \in S \setminus S^{0}} \prod_{i \in S^{j}} P_{ij}^{\text{DR}}
= \prod_{j \in S \setminus S^{0}} \prod_{i \in S^{j}}
\frac{y_{ij}\phi^{u_j}}{\sum_{l=1}^{N} y_{il} \phi^{u_l}},
\end{equation}
where $\boldsymbol{R}$ is a vector encoding the recruitment order, and $\boldsymbol{D}_k^{u}$ denote the vector whose $i$-th entry equals the number of ties that individual $i$ has to individuals with $u = k$, for $k \in \{0,1\}$. The set $S \setminus S^{0}$ corresponds to all sampled individuals excluding the seeds. For each recruited individual $j$, the set $S^{j} = \{i \in S : S_{ij} = 1 \}$ identifies the unique recruiter of $j$. Under the assumption that recruitment events occur independently, the likelihood function in Equation~\eqref{eq:likdr} corresponds to the probability of the observed recruitment structure. The seeds are excluded from this likelihood function since they are not recruited by other participants. 

Once $\widehat{\phi}$ is estimated, it is substituted into Equation~\eqref{eq:drstat} to calculate the unnormalized stationary distribution, $\hat{\pi}_{i}^{\text{DR}}~\propto~\widehat{\phi}^{u_{i}}(\widehat{\phi} d_{i1}^{u} + d_{i0}^{u})$. These weights serve as the estimated sampling probabilities ($\hat{p}_{i}$), up to a constant of proportionality, within the extended prevalence estimators. For the extended Volz-Heckathorn (VH) estimator, the authors obtained:
\begin{equation}
  \hat{\mu}_{DR}^{II} =  \left[\displaystyle\displaystyle\sum_{i=1}^{N}\displaystyle\frac{S_{i}z_{i}}{\hat{\phi}^{u_{i}}(\hat{\phi} d_{i1}^{u} + d_{i0}^{u})}\right]
\left[\displaystyle\displaystyle\sum_{i=1}^{N}\displaystyle\frac{S_{i}}{\hat{\phi}^{u_{i}}(\hat{\phi} d_{i1}^{u} + d_{i0}^{u})}\right]^{-1}. 
\label{eq:VH2}
\end{equation}

Similarly, their extended Lu estimator, $\hat{\mu}_{DR}^{ego}$, retains the same structure as Equation~\eqref{eq:mu}. In this case, the quantities $C_{k,1-k}$ and $D_k$ for $k \in \{0,1\}$ are also estimated using a Hájek-type estimator, but using the weights $\hat{p}_i$:
\begin{flalign}
 \widehat{C}^{ego.DR}_{k,1-k} &=\left[\sum_{i=1}^{N}\frac{S_{i}d_{i(1-k)}^{z}\mathbbm{1}[z_{i} = k]}{\widehat{\phi}^{u_{i}}(\widehat{\phi} d_{i1}^{u} + d_{i0}^{u})}\right]\left[\sum_{i=1}^{N}
\frac{S_{i}d_i\mathbbm{1}[z_{i}=k]}{\widehat{\phi}^{u_{i}}(\widehat{\phi} d_{i1}^{u} + d_{i0}^{u})}
\right]^{-1},  \text{ and} \\[6pt]
\widehat{D}_{k}^{ego.DR} &= 
\left[\displaystyle\sum_{i=1}^{N}
\frac{S_{i}d_{i}\mathbbm{1}{[z_{i} = k]}}{\widehat{\phi}^{u_{i}}(\widehat{\phi} d_{i1}^{u} + d_{i0}^{u})}
\right]
\left[\displaystyle\sum_{i=1}^{N}
\frac{S_{i}\mathbbm{1}{[z_{i} = k]}}{\widehat{\phi}^{u_{i}}(\widehat{\phi} d_{i1}^{u} + d_{i0}^{u})}\right]^{-1}.
\end{flalign}

They also proposed an extended version of the SH estimator, denoted $\hat{\mu}_{DR}^{I}$. However, they showed that the performance was generally inferior to that of $\hat{\mu}_{Lu}^{ego}$ and $\hat{\mu}_{DR}^{ego}$ under differential recruitment \citep{Beaudry2020}. Consequently, we exclude $\hat{\mu}_{SH}^{I}$ and $\hat{\mu}_{DR}^{I}$ from the remainder of our analysis.

\subsection{Uncertainty of estimators} \label{sec:bootlit}

In this section, we review several bootstrap methods for variance estimation in RDS. These include the procedures of \citet{Salganik2006} for the SH and VH estimators, \citet{Lu2013} for the Lu estimator, and \citet{Beaudry2020} for their DR estimators, as well as the neighborhood bootstrap of \citet{Yauck2022}. 

One of the primary objectives of these bootstrap procedures is to preserve the dependence structure characteristic of RDS data. Each method attempts to do so through its resampling scheme. Aside from this difference, all approaches follow the same general principle: generating $B$ replicate samples from the observed RDS data to estimate variances and construct confidence intervals. Below, we outline the resampling steps for each procedure.

In the Salganik bootstrap method \citep{Salganik2006}, participants are partitioned into two sets, $R_{0}$ and $R_{1}$, according to the infection status of their recruiter in the sample. Specifically, $R_{0}$ ($R_{1}$) contains individuals recruited from participants with infection status $0$ ($1$). The procedure begins by randomly selecting a seed from the sample. If the seed has infection status $0$ ($1$), the next individual is sampled at random from $R_{0}$ ($R_{1}$). Recruitment then proceeds iteratively: for each of the remaining $n-1$ steps, conditional on the infection status $z_i$ of the current recruiter $i$, the next participant is sampled with replacement at random from the corresponding set $R_{z_i}$. This procedure aims to mimic the recruitment process while preserving the observed dependence of recruitment on infection status. 

In the Lu bootstrap method \citep{Lu2013}, resampling also proceeds sequentially. A node is first selected at random from the RDS sample. Subsequent individuals are then sampled with replacement according to the $2 \times 2$ recruitment transition matrix, with entries:
\begin{equation} 
\frac{\widehat{C}_{kl}^{ego}}{n_l}, \quad  k, l \in \{0,1\},
\end{equation}
where $k$ and $l$ denote the infection status of the recruiter and candidate node, respectively, and where $n_l = \sum_{i=1}^{N} S_i \mathbbm{1}[z_i = l]$ is the number of sampled nodes with infection status $l$. The value $\widehat{C}_{kl}^{ego}$ is divided by $n_l$ because it represents a transition probability between infection statuses, whereas the bootstrap resampling is conducted at the node level. This process is repeated until a replicate sample of size $n$ is obtained.

To estimate the variance of $\hat{\mu}_{DR}^{I}$, $\hat{\mu}_{DR}^{II}$, and $\hat{\mu}_{DR}^{ego}$, \citet{Beaudry2020} proposed two variations of the bootstrap approach of \citet{Lu2013}. In this paper, we describe and consider only the first variation. The original Lu bootstrap uses a transition matrix based on cross-group tie proportions ($\widehat{C}_{kl}^{ego}$). A direct analogue in the DR setting, replacing $\widehat{C}_{kl}^{ego}$ with $\widehat{C}_{kl}^{ego.DR}$, would be inappropriate, since $\widehat{C}_{kl}^{ego.DR}$ estimates cross-group tie proportions rather than recruitment behavior. To capture the recruitment dynamics under DR, the authors construct a $2 \times 2$ transition matrix partitioned by the characteristic $u$, with entries given by Hájek-style estimates of between-group recruitment proportions.

Specifically, nodes are grouped as $\mathcal{U}_k = \{i : u_i = k\}$ for $k \in \{0,1\}$. The probability that a node $i \in \mathcal{U}_{u_i}$ transitions to a node $j \in \mathcal{U}_{u_j}$ depends both on the number of observed cross-group ties between $\mathcal{U}_{u_i}$ and $\mathcal{U}_{u_j}$ and on the estimated differential recruitment propensity $\hat{\phi}$.

The neighborhood bootstrap (NB) methodology \citep{Yauck2022} differs from the approaches described above in that it does not rely on a two-state MC. It was originally designed to assess the uncertainty of the VH estimator $\hat{\mu}_{VH}^{II}$. The resampling procedure is based on the recruitment tree and proceeds in two main steps. First, $r$ recruiters are randomly selected with replacement, where $r$ is the total number of recruiters in the original sample. Second, all immediate recruits, the {\it children}, of the selected recruiters are included. The final bootstrap replicate is obtained by taking the union of the selected recruiters and their recruits. Note that this approach does not guarantee that each replicate will have the same size as the original RDS sample.

The primary purpose of generating $B$ bootstrap replicates is to estimate the variability of the prevalence estimators and construct corresponding confidence intervals. For each replicate, the prevalence estimate $\hat{\mu}^*_{b}$ is computed, where~${}^*$ denotes the estimator of interest as described in Section~\ref{sec:prevalence}, and $b = 1,\ldots,B$. The methodologies discussed in this section employ different approaches to construct confidence intervals. Specifically, the Salganik, Lu, and NB bootstraps use the percentile method, whereas the DR bootstrap relies on a studentized interval.

\section{Multivariate differential recruitment}\label{sec:MDR}

Thus far, we have discussed estimators under both random recruitment and differential recruitment (DR), along with their corresponding variance estimators. Estimators that assume random recruitment fail to account for biases introduced by participants' recruitment behavior, and may therefore be substantially biased in the presence of DR \citep{Gile2015, Shi2019}. DR estimators have been shown to improve prevalence estimation under non-random recruitment substantially, but are designed to address DR for only a single categorical variable \citep{Beaudry2020}. In practice, recruitment patterns may be influenced by multiple continuous or categorical variables simultaneously. In this section, we introduce a mathematical framework for Multivariate Differential Recruitment (MDR), along with extended prevalence estimators and their associated variance estimation methodology.

\subsection{Sampling model under MDR} \label{sec:MDRmodel}

MDR occurs in RDS when multiple factors simultaneously influence participants' recruitment behavior. For example, age may affect recruitment, as respondents may be more likely to recruit peers of a similar age. Other factors, such as the proximity between respondents' workplaces and residences or the nature of their relationship (e.g., close friendship versus acquaintance), may also shape recruitment patterns. The proposed sampling model for RDS accommodates the influence of multiple variables on recruitment. 

We model the RDS sampling process as a Markov chain (MC) with the nodes of the underlying network as its state space. The MC captures MDR through transition probabilities that are functions of node- and tie-specific features, and unknown parameters that weight the influence of each variable on the recruitment process. Specifically, let $\boldsymbol{x}_{ij}$ be a $K$-dimensional vector of covariates that may influence recruitment patterns between nodes $i$ and $j$ in the target population. For a fixed pair $(i,j)$, we define:
\begin{equation}
\boldsymbol{x}_{ij} = (r_{j1}, \ldots, r_{jK_{1}}, w_{ij1}, \ldots, w_{ijK_{2}})^{\top},
\end{equation}

where $K = K_{1}+K_{2}$. Under this notation, $r_{jk_{1}}$ (for $k_1 = 1, \ldots, K_1$) denotes the $k_1$-th covariate describing node $j$, such as age or nationality. Similarly,  $w_{ijk_{2}} = w_{jik_{2}}$ (for $k_2 = 1, \ldots, K_2$) represents the $k_2$-th symmetric attribute of the relationship between nodes $i$ and $j$, such as their weekly meeting frequency. We emphasize that the covariates in $\boldsymbol{x}_{ij}$ may be either continuous or categorical. In addition, we consider a vector of unknown coefficients $\boldsymbol{\beta} = (\beta_1, \ldots, \beta_K)^{\top}$ which reflects the weight of each covariate in the vector $\boldsymbol{x}_{ij}$. Under this parameterization, the transition probabilities of the MC modeling MDR are given by:
\begin{equation}\label{eq:mcextended}
    P_{ij}^{\text{MDR}} = \frac{y_{ij} e^{\boldsymbol{x}_{ij}^{\top}\boldsymbol{\beta}}}{\sum_{l=1}^{N}y_{il} e^{\boldsymbol{x}_{il}^{\top}\boldsymbol{\beta}}} \quad\quad i,j = 1,\ldots,N.
\end{equation}

\begin{figure}[htbp]
   \centering 
\begin{tikzpicture}[node distance=2.5cm, auto, >=stealth, bend angle=20]

    \tikzstyle{person}=[circle, draw, thick, minimum size=1.2cm, fill=blue!10]
   \tikzstyle{attr}=[font=\small\itshape, text=gray!80!black]
    \tikzstyle{math_label}=[font=\small, text=blue!70!black]

    \node[person,fill=green!40] (A) at (0,0) {\textbf{A}};
    \node[person] (B) at (4.5,1.85) {\textbf{B}};
    \node[person] (C) at (4.5,-1.85) {\textbf{C}};

    \node[attr, anchor=west,text=black] at (5.2, 2.2) {$\mathrm{edu}_B=1$};
    \node[attr, anchor=west,text=black] at (5.2, 1.8) {$w_{AB} \hspace{.07cm}= 5$};

    \node[attr, anchor=west,text=black] at (5.2, -1.7) {$\mathrm{edu}_C=0$};
    \node[attr, anchor=west,text=black] at (5.2, -2.1) {$w_{AC} \hspace{.07cm}= 10$};

    \draw[->, line width=2.4pt, blue!60] (A) -- (B) 
        node[midway, above, sloped, math_label, black] {$P_{AB}^{\text{MDR}} \propto e^{0.5(1) - 0.14(5)}$};
    
    \draw[->, line width=0.8pt, blue!60] (A) -- (C) 
       node[midway, above, sloped, math_label, black] {$P_{AC}^{\text{MDR}} \propto e^{0.5(0) - 0.14(10)}$};

    \node[draw, dashed, inner sep=5pt, rounded corners] at (11,1.9) {
        $\displaystyle\frac{P_{AB}^{\text{MDR}}}{P_{AC}^{\text{MDR}}} = \frac{e^{0.5(1) - 0.14(5)}}{e^{0.5(0) - 0.14(10)}} \approx 3.3$
    };
\end{tikzpicture}
\caption{Comparison of recruitment probabilities under the MDR model with two variables (edu and $w$) given $\boldsymbol{\beta} = (0.5, -0.14)$. The thickness of the ties indicates that node $A$ recruits node $B$ with approximately 3.3 times the probability of recruiting $C$.}
\label{fig:mdr_logic}
\end{figure}
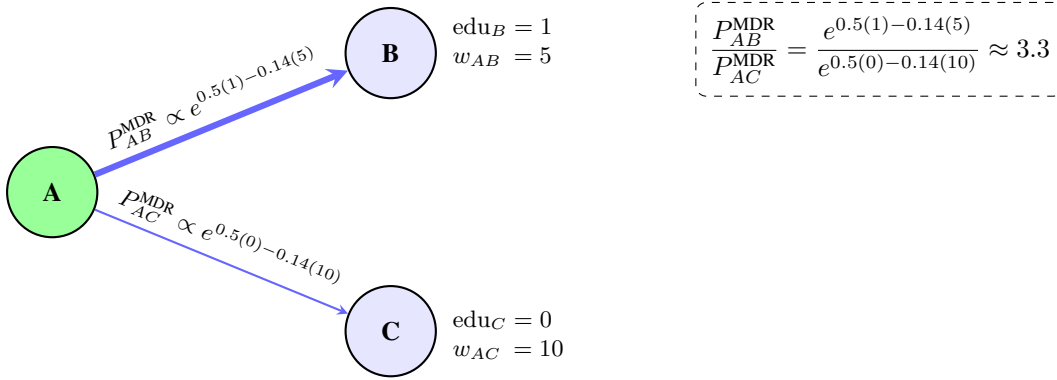

Figure~\ref{fig:mdr_logic} illustrates how multiple covariates induce MDR in a simple network where recruiter $A$ is connected to potential recruits $B$ and $C$. Recruitment depends on two covariates: (i) a node-specific attribute, a binary education level variable (e.g.,~$\text{edu}_B = 1$ and $\text{edu}_C = 0$), and (ii) a tie-specific attribute, the absolute age difference, $w_{ij} = |\text{age}_i - \text{age}_j|$ (e.g.,~$w_{AB} = 5$ and $w_{AC} = 10$). Suppose the regression coefficients are $\beta_1 = 0.5$ and $\beta_2 = -0.14$. The negative value of $\beta_2$ indicates that recruitment probability increases as the age gap narrows. In this scenario, node $A$ is approximately 3.3 times more likely to recruit $B$ than $C$, as shown by the ratio of their transition probabilities.

Having defined the MDR MC transition probabilities, our next task is to derive the stationary distribution of the Markov process. This distribution will serve as the basis for estimating the RDS sampling probabilities used in the extended prevalence estimators. Result~\ref{res:piMDR} establishes the existence of a unique stationary distribution and specifies the conditions under which it holds, followed by its proof.

\begin{result}\label{res:piMDR}
Consider a Markov chain on the state space of nodes in a fully connected undirected network without self-ties, with transition probabilities:
\begin{equation}
P_{ij}^{\text{MDR}} = \frac{y_{ij}\, e^{\boldsymbol{x}_{ij}^{\top}\boldsymbol{\beta}}}{\sum_{l=1}^{N} y_{il} e^{\boldsymbol{x}_{il}^{\top}\boldsymbol{\beta}}}, 
\quad i,j = 1,\ldots,N.
\end{equation}

Then the chain admits a unique stationary distribution given by:
\begin{equation}
    \pi_{i}^{\text{MDR}} = \frac{1}{\kappa} \sum_{j=1}^{N} y_{ij}\, e^{\boldsymbol{x}_{ij}^{\top}\boldsymbol{\beta} + \boldsymbol{r}_{i}^{\top}\boldsymbol{\alpha}},
    \quad i = 1,\ldots,N, \label{eq:statdist}
\end{equation}
where $\boldsymbol{\alpha} = (\beta_{1},\ldots,\beta_{K_{1}})^{\top}$ is the subvector of $\boldsymbol{\beta}$ corresponding to node-specific covariates 
$\boldsymbol{r}_{i} = (r_{i1},\ldots,r_{iK_{1}})^{\top}$, and the normalizing constant $\kappa = \sum_{l=1}^{N} \sum_{j=1}^{N} y_{lj}\, e^{\boldsymbol{x}_{lj}^{\top}\boldsymbol{\beta} + \boldsymbol{r}_{l}^{\top}\boldsymbol{\alpha}}$ ensures $\sum_{i=1}^{N} \pi_{i}^{\text{MDR}} = 1$.
\end{result}

\textbf{Proof:}
Let $\boldsymbol{\gamma}=(\beta_{K_{1}+1},\ldots,\beta_{K})^{\top}$, $\boldsymbol{w}_{ij}=\boldsymbol{w}_{ji}=(w_{ij1},\ldots,w_{ijK_{2}})^{\top}$. Then, the balanced equations are:
\begin{flalign}
\begin{split}
\sum_{i=1}^{N}P_{ij}^{\text{MDR}}\pi_{i}^{\text{MDR}}  &= \sum_{i=1}^{N}\left[\frac{y_{ij} e^{\boldsymbol{x}_{ij}^{\top}\boldsymbol{\beta}}}{\sum_{l=1}^{N}y_{il} e^{\boldsymbol{x}_{il}^{\top}\boldsymbol{\beta}}}\right]\left[\frac{\sum_{l=1}^{N}y_{il} e^{\boldsymbol{x}_{il}^{\top}\boldsymbol{\beta}+\boldsymbol{r}_{i}^{\top}\boldsymbol{\alpha}}}{\kappa}\right] \notag\\
    &= \dfrac{1}{\kappa}\sum_{i=1}^{N}y_{ij}e^{\boldsymbol{x}_{ij}^{\top}\boldsymbol{\beta}+\boldsymbol{r}_{i}^{\top}\boldsymbol{\alpha}} 
    = \dfrac{1}{\kappa}\sum_{i=1}^{N}y_{ij}e^{(\boldsymbol{r}_{j}^{\top}\boldsymbol{\alpha}+\boldsymbol{w}_{ij}^{\top}\boldsymbol{\gamma}) + \boldsymbol{r}_{i}^{\top}\boldsymbol{\alpha}} \hspace{1cm} \left(\text{since } e^{\boldsymbol{x}_{ij}^{\top}\boldsymbol{\beta}} = e^{(\boldsymbol{r}_{j}^{\top}\boldsymbol{\alpha}+\boldsymbol{w}_{ij}^{\top}\boldsymbol{\gamma})}\right)\\
     &= \dfrac{1}{\kappa}\sum_{i=1}^{N}y_{ji} e^{(\boldsymbol{r}_{i}^{\top}\boldsymbol{\alpha}+\boldsymbol{w}_{ji}^{\top}\boldsymbol{\gamma}) + \boldsymbol{r}_{j}^{\top}\boldsymbol{\alpha}}
    = \dfrac{1}{\kappa}\sum_{i=1}^{N}y_{ji} e^{\boldsymbol{x}_{ji}^{\top}\boldsymbol{\beta} + \boldsymbol{r}_{j}^{\top}\boldsymbol{\alpha}}
    = \pi_{j}^{\text{MDR}}.
\end{split}
\end{flalign}

Therefore, we have shown that $\sum_{i=1}^{N}P_{ij}^{\text{MDR}}\pi_{i}^{\text{MDR}} = \pi_{j}^{\text{MDR}} ~ \forall~ j \in \{1,\ldots,N\}. \quad \square$

The stationary distribution, which we assume determines the sampling probabilities, is a function of both the node and its local network characteristics, as well as the unknown parameters $\boldsymbol{\beta}$. Following \cite{Beaudry2020}, we estimate $\boldsymbol{\beta}$ by maximizing the likelihood function:
\begin{equation}
    \mathcal{L}(\boldsymbol{\beta} \mid \boldsymbol{R}, \boldsymbol{Y}^{obs}, \boldsymbol{X}^{obs}) = \displaystyle\prod_{j \in S \setminus S^{0}} \prod_{i \in S^{j}} P_{ij}^{\text{MDR}} = \displaystyle\prod_{j \in S \setminus S^{0}} \prod_{i \in S^{j}}\frac{y_{ij}e^{\boldsymbol{x}_{ij}^{\top}\boldsymbol{\beta}}}{\sum_{l=1}^{N}y_{il}e^{\boldsymbol{x}_{il}^{\top}\boldsymbol{\beta}}},
\label{eq:likelihood}
\end{equation}
where $\boldsymbol{R}$ is a vector encoding the recruitment order, and $\boldsymbol{Y}^{obs}$ and $\boldsymbol{X}^{obs}$ denote the observed portion of the network and MDR covariates, respectively. Consistent with Equation~\eqref{eq:likdr}, let $S \setminus S^{0}$ denote the set of all non-seed respondents. For each individual~$j \in S \setminus S^{0}$, we define $S^{j} = \{ i \in S : S_{ij} = 1 \}$ as the set identifying the unique recruiter of node~$j$.
Under the assumption that recruitment events occur independently, Equation~\eqref{eq:likelihood} characterizes the probability of observing the recruitment structure. The resulting likelihood function is maximized using numerical optimization techniques. In particular, optimization is performed using the \texttt{optim} function implemented in the R statistical software \citep{R-base}.

\subsection{Extended estimators with MDR} \label{sec:MDRprevalence}

In this section, we extend the DR estimators $\hat{\mu}_{DR}^{II}$ and $\hat{\mu}_{DR}^{ego}$ to their MDR counterparts, $\hat{\mu}_{MDR}^{II}$ and $\hat{\mu}_{MDR}^{ego}$. Their construction parallels that of Section~\ref{sec:DRprev}, with the difference that the sampling probabilities are now given by $\hat{\pi}_{i}^{\text{MDR}}$ instead of $\hat{\pi}_{i}^{\text{DR}}$. These probabilities are obtained by substituting the parameter estimates from the maximization of Equation~\eqref{eq:likelihood} into the stationary distribution in Equation~\eqref{eq:statdist}. Up to a normalizing constant, they take the form $\hat{\pi}_{i}^{\text{MDR}} \propto \sum_{j=1}^{N} y_{ij} e^{\boldsymbol{x}_{ij}^{\top}\hat{\boldsymbol{\beta}}+\boldsymbol{r}_{i}^{\top}\hat{\boldsymbol{\alpha}}}, i=1,\ldots,N.$ Therefore, $\hat{\mu}^{II}_{MDR}$ is given by:
\begin{equation}
  \hat{\mu}_{MDR}^{II} =  \left[\displaystyle\sum_{i=1}^{N}\frac{S_{i}z_{i}}{\sum_{j=1}^{N}y_{ij}e^{\boldsymbol{x}_{ij}^{\top}\hat{\boldsymbol{\beta}}+\boldsymbol{r}_{i}^{\top}\hat{\boldsymbol{\alpha}}}}\right]
\left[\displaystyle\sum_{i=1}^{N}\frac{S_{i}}{\sum_{j=1}^{N}y_{ij}e^{\boldsymbol{x}_{ij}^{\top}\hat{\boldsymbol{\beta}}+\boldsymbol{r}_{i}^{\top}\hat{\boldsymbol{\alpha}}}}\right]^{-1}. 
\label{eq:VH3}
\end{equation}

The MDR-based estimator $\hat{\mu}_{MDR}^{ego}$ follows the same general form as Equation~\eqref{eq:mu}. In this case, the components $C_{k,1-k}$ and $D_k$ for $k \in \{0,1\}$ are estimated using H{\'a}jek-type estimators that incorporate the MDR-adjusted sampling probabilities, $\hat{\pi}^{\text{MDR}}_i$, rather than the DR-based ones:
\begin{flalign}
\label{eq:ego_mdr1}
 \widehat{C}^{ego.MDR}_{k,1-k} &=\left[\sum_{i=1}^{N}\frac{S_{i}d_{i(1-k)}^{z}\mathbbm{1}[z_{i} = k]}{\sum_{j=1}^{N} y_{ij} e^{\boldsymbol{x}_{ij}^{\top}\hat{\boldsymbol{\beta}}+\boldsymbol{r}_{i}^{\top}\hat{\boldsymbol{\alpha}}}}\right]\left[\sum_{i=1}^{N}
\frac{S_{i}d_i\mathbbm{1}[z_{i}=k]}{\sum_{j=1}^{N} y_{ij} e^{\boldsymbol{x}_{ij}^{\top}\hat{\boldsymbol{\beta}}+\boldsymbol{r}_{i}^{\top}\hat{\boldsymbol{\alpha}}}}
\right]^{-1},  \text{ and} \\[6pt]
\label{eq:ego_mdr2}
\widehat{D}_{k}^{ego.MDR} &= 
\left[\displaystyle\sum_{i=1}^{N}
\frac{S_{i}d_{i}\mathbbm{1}{[z_{i} = k]}}{\sum_{j=1}^{N} y_{ij} e^{\boldsymbol{x}_{ij}^{\top}\hat{\boldsymbol{\beta}}+\boldsymbol{r}_{i}^{\top}\hat{\boldsymbol{\alpha}}}}
\right]
\left[\displaystyle\sum_{i=1}^{N}
\frac{S_{i}\mathbbm{1}{[z_{i} = k]}}{\sum_{j=1}^{N} y_{ij} e^{\boldsymbol{x}_{ij}^{\top}\hat{\boldsymbol{\beta}}+\boldsymbol{r}_{i}^{\top}\hat{\boldsymbol{\alpha}}}}\right]^{-1}.
\end{flalign}

\subsection{Uncertainty of estimators} \label{sec:bootnb}

We adopt the NB procedure to estimate the variance of $\hat{\mu}^{II}_{MDR}$ and $\hat{\mu}^{ego}_{MDR}$. Although originally developed for $\hat{\mu}^{II}_{VH}$, we apply this methodology to our estimators because its resampling of recruiter-recruit groups preserves the recruitment dependencies inherent in the MDR process. To implement this, we slightly modify the algorithm described in Section~\ref{sec:bootlit} to ensure each bootstrap replicate size remains constant and equal to the original sample size $n$.

The key idea is that, since respondents can invite at most $c$ contacts, if all $r$ recruiters distribute all $c$ coupons, the resulting replicate would contain $r(1+c)$ individuals. Thus, by selecting $r_e = \lceil n/(1+c) \rceil$ recruiters and including their immediate recruits, we obtain a replicate approximately of size $n$. In practice, however, not all recruiters distribute all their coupons, so the replicate size $n_b$ may still differ from $n$. To solve this issue, we introduce additional conditions to adjust the replicate until $n_b = n$. Algorithm \ref{alg1} describes each step of our modified NB procedure.

\begin{algorithm}[ht!]
\SetKwInOut{Input}{input}
\SetKwInOut{Output}{output}
 \caption{Modified NB procedure for fixed sample size.} \label{alg1}
\SetAlgoLined
\KwData{$c, n, r_{e}, \mathcal{R}$ (set of all recruiters).}
Select, completely at random and with replacement, $r_{e}$ recruiters from $\mathcal{R}$, set them as $\mathcal{R}_b$.\\
Set $\mathcal{C}_{\mathcal{R}_{b}} = \{j : j ~ \text{is a immediate children of someone in} ~ \mathcal{R}_{b}\}$.\\
Compute $n_{b} = |\mathcal{R}_{b} \cup \mathcal{C}_{\mathcal{R}_{b}}|$ as the sample size of the initial $b$-replicate.\\
Compute $\delta_{b} = n_{b}-n$.\\
\While{$(\delta_{b} < 0)$}{
    -- Compute $k = \lceil{|\delta_{b}|/(1+c)\rceil{}}$. \\
    -- Select completely at random and with replacement $k$ new recruiters from $\mathcal{R}$. Set this new recruiters as $\widetilde{\mathcal{R}}_{b}$. \\
    -- Select the immediate children of the new recruiters in $\widetilde{\mathcal{R}}_{b}$. Set them as $\widetilde{\mathcal{C}}_{\mathcal{R}_{b}}$. \\
    -- Update $\mathcal{R}_{b} = \mathcal{R}_{b} \cup \widetilde{\mathcal{R}}_{b}$ and $\mathcal{C}_{\mathcal{R}_{b}} = \mathcal{C}_{\mathcal{R}_{b}} \cup \widetilde{\mathcal{C}}_{\mathcal{R}_{b}}$. \\
    -- Compute $n_{b}$ and $\delta_{b}$ as established in step 3 and 4 respectively. 
}
\If{$(\delta_{b} > 0)$}{
    -- Remove $\delta_{b}$ excess nodes. 
}
\KwResult{$b$-replicate.}

\end{algorithm}
 
As seen in Algorithm \ref{alg1} (line 13), $\delta_{b}$ nodes are sometimes removed from the replicate. The excess nodes are eliminated by randomly selecting $\delta_{b}$ nodes such that each recruiter in the replica keeps at least one recruit. This prevents the creation of isolated recruiters who, despite remaining in the bootstrap replicate, would fail to contribute to the likelihood function due to the absence of recruitment ties. Practically, this is achieved by either removing an entire recruitment cluster of $\delta_{b}$ size (a recruiter and all their recruits) or by pruning $\delta_{b}$ individual recruits, provided their removal does not leave a recruiter with zero ties.

The modified NB procedure is repeated $B$ times. For each replicate, the prevalence estimates $\hat{\mu}^{II}_{MDR}$ and $\hat{\mu}^{ego}_{MDR}$ are computed, and the standard deviation of the resulting $B$ estimates is used as the bootstrap standard error. As discussed in Section~\ref{sec:bootlit}, the original NB method constructs confidence intervals using the percentile approach. In contrast, we construct confidence intervals based on the normal approximation using standard normal quantiles. This choice ensures a fair comparison across all bootstrap methodologies considered in the simulation study.

\section{Simulation study} \label{sec:4}

This section presents a simulation study designed to evaluate the performance of the methodology proposed in Section~\ref{sec:MDR}. Specifically, we examine the behavior of the proposed estimators both in the presence and absence of Multivariate Differential Recruitment (MDR) and compare them to the alternative estimators described in Section~\ref{sec:prevalence}. The simulation framework explores a range of sampling strategies and network properties to provide a thorough assessment of estimator accuracy across diverse scenarios.

\subsection{Populations}

We conducted a simulation study in which 15 social networks were generated per scenario, each with $N = 1000$ individuals. Individual ages were independently drawn from a Gamma distribution with mean 26 and standard deviation $\sqrt{26} \approx 5.1$ (i.e., $X_i \sim \operatorname{Gamma}(26,1)$). Conditional on age, infection status $Z_i$ was then simulated independently for each individual \citep{Banze2024} according to the following binary model:
\begin{equation}
\label{eq:zdadox}
 Z_{i} \mid x_{i} \sim \operatorname{Bin}(1, \mu_{i}), \quad \text{with} \quad \text{logit}\left(\mu_i\right) = -4.0 + 0.09 x_{i},
\end{equation}

\noindent where $x_i$ denotes the realized age of individual $i$. The coefficients ($-4.0$ and $0.09$) imply an expected prevalence of 0.1 at age 20
and 0.4 at age 40, 
capturing the positive association between age and prevalence that we aimed to model. For each network, we also computed the true prevalence as $\mu = \sum_{i=1}^{N} z_i / N$, with the resulting average prevalence ranging from 0.16 to 0.17 across networks.

To generate networks, we modeled each adjacency matrix $\boldsymbol{Y}$ using an Exponential Random Graph Model (ERGM) \citep{Frank1986,Hunter2006}. ERGMs specify a probability distribution over the space of networks and allow dependence on both covariates (e.g., nodal and dyadic) and structural network features (e.g.,~network density). In practice, many factors may influence tie formation \citep{Lusher2013}. In our specification, we included age difference as a dyadic predictor to incorporate a continuous covariate and to reflect the well-documented role of age in social interactions \citep{McPherson2001}. This term induces age-based homophily, favoring ties between individuals of similar age \citep{Degenne1999, Currarini2009}. Specifically, symmetric $\boldsymbol{Y}$ matrices were generated according to:
\begin{equation}
    \operatorname{P}(\boldsymbol{Y} = \boldsymbol{y} \mid \boldsymbol{x}, \boldsymbol{\eta}) =  \frac{1}{C(\boldsymbol{\eta} \mid \boldsymbol{x})}\exp\left\{\eta_{1}\sum_{i < j}y_{ij} + \eta_{2}\sum_{i < j}y_{ij}|x_{i}-x_{j}|\right\},
    \label{eq:ergm1}
\end{equation}

\noindent where $\boldsymbol{\eta} = (\eta_{1},\eta_{2})^{\top}$ denotes the unknown ERGM parameters, and $C(\boldsymbol{\eta} \mid \boldsymbol{x})$ is the normalizing constant. Our specification is equivalent to assuming that each potential tie $Y_{ij}$ is drawn independently from a Bernoulli distribution, where the probability of a tie between nodes $i$ and $j$ is given by:
\begin{equation}\label{eq:ergm}
    \operatorname{P}(Y_{ij} = 1 \mid x_{i}, x_{j}) = \frac{e^{(\eta_{1}+\eta_{2}|x_{i}-x_{j}|)}}{1+e^{(\eta_{1}+\eta_{2}|x_{i}-x_{j}|)}}.
\end{equation}

The parameter $\eta_2$ in Equations~\eqref{eq:ergm1} and \eqref{eq:ergm} governs the strength of age-based homophily. Since $\eta_2$ is defined on the log-odds scale, its direct interpretation in terms of tie probabilities is not straightforward. To facilitate interpretation, we introduce the parameter $\tau$, defined as follows:
\begin{equation}
\label{eq:tau}
\tau = \frac{\operatorname{P}(Y_{ij}=1 \mid |x_{i}-x_{j}| \leq 5)}{\operatorname{P}(Y_{ij}=1 \mid |x_{i}-x_{j}| > 5)}.
\end{equation}

The parameter $\tau$ represents the relative propensity for ties between individuals whose ages differ by at most 5 years compared with those whose age difference exceeds 5 years. This age-similarity threshold is motivated by the fact that $\boldsymbol{X} \sim \operatorname{Gamma}(26,1)$ yields a median absolute age difference of approximately 5 years. Networks were generated under three levels of homophily, corresponding to $\tau \in \{1.0,3.2,5.1\}$ (none, moderate, and high homophily, respectively). 
We selected $\boldsymbol{\eta}$ to obtain those values of $\tau$ while maintaining an average degree of $\bar{d}=12$.

We verified that each simulated network contained no isolated nodes. Because stronger homophily increases the chance of generating disconnected networks, a small number of realizations were discarded. This had only a minor effect on the overall prevalence, which remained approximately 0.16 to 0.17 across scenarios.

\subsection{Sampling}

To assess the impact of MDR on RDS estimation, samples were generated from simulated networks under scenarios ranging from no MDR to high MDR. In these scenarios, MDR was induced by four covariates: the potential recruit's age ($x_j$), infection status ($z_j$), the interaction term ($x_j z_j$), and the absolute age difference between recruiter $i$ and potential recruit $j$ ($|x_i-x_j|$). The first three covariates represent characteristics of the potential recruit (node $j$), while the last is a symmetric attribute of the dyad $(i,j)$. The vector $\boldsymbol{x}_{ij}$ contains these covariates and is weighted by the parameter vector $\boldsymbol{\beta} = (\beta_{1}, \beta_{2}, \beta_{3}, \beta_{4})^{ \top}$, which was varied across simulation settings to generate different levels of recruitment bias.

Because the $\boldsymbol{\beta}$-parameters may differ in scale across covariates, and following ideas similar to those in \cite{Beaudry2020}, a summary metric $\phi$ was introduced to capture the overall strength of MDR within a network. Specifically, $\phi^{\text{MDR}}$ is the average of node-specific $\phi_i^{\text{MDR}}$, defined as:  
\begin{equation}
\phi_i^{\text{MDR}} = 
\begin{cases}
\left[\displaystyle\sum_{k,j \in \mathcal{N}_i,\, k \ne j}  \frac{P_{ik}^{\text{MDR}}}{P_{ij}^{\text{MDR}}} \, \mathbbm{1} \left[ \frac{P_{ik}^{\text{MDR}}}{P_{ij}^{\text{MDR}}} \geq 1 \right]\right]\left[\displaystyle\sum_{k,j \in \mathcal{N}_i,\, k \ne j}\mathbbm{1}\left[ \frac{P_{ik}^{\text{MDR}}}{P_{ij}^{\text{MDR}}} \geq 1 \right]\right]^{-1}, & \text{if } d_i > 1 \\[21pt]
1, & \text{if } d_i = 1,
\end{cases}
\end{equation}
where $\mathcal{N}_{i} = \{l : y_{il} = 1\}$ is the set of contacts or neighbors of node $i$. Intuitively, the metric $\phi_i^{\text{MDR}}$ summarizes the extent to which recruitment probabilities for node $i$ vary across its alters. Values greater than one indicate recruitment imbalance, whereas $\phi_i^{\text{MDR}}=1$ corresponds to random recruitment with equal transition probabilities across alters. For example, if node $i$ is connected to nodes $A$, $B$, and $C$, with transition probabilities $P_{iA}^{\text{MDR}}=0.60$, $P_{iB}^{\text{MDR}}=0.30$, and $P_{iC}^{\text{MDR}}=0.10$, the ratios greater than or equal to one are $P_{iA}^{\text{MDR}}/P_{iB}^{\text{MDR}}=2$, $P_{iA}^{\text{MDR}}/P_{iC}^{\text{MDR}}=6$, and $P_{iB}^{\text{MDR}}/P_{iC}^{\text{MDR}}=3$. Therefore, $\phi_i^{\text{MDR}}=(2+6+3)/3 \approx 3.67$, indicating a substantial imbalance in recruitment probabilities across alters. The overall MDR strength, $\phi^{\text{MDR}}$, is defined as the average of $\phi_i^{\text{MDR}}$ across all individuals in the network.

Table~\ref{tab:sc} summarizes the nine simulation scenarios, which were obtained by crossing three levels of network homophily, $\tau \in \{1.0,3.2,5.1\}$ (none, moderate, and high), with three levels of MDR strength (none, moderate, and high).

\begin{table}[ht!]
\centering
\caption{Parameters summarizing the overall strength of the network homophily ($\tau$) and MDR ($\phi^{\text{MDR}}$) in the simulation study scenarios.}
\begin{tabular}{ccccc}
\cmidrule{2-5}
&\multirow{2}{*}{$(\tau$, $\phi^{\text{MDR}})$} & \multicolumn{3}{c}{MDR} \\
\cmidrule{3-5}
& &  \text{None} &   \text{Moderate} &  \text{High} \\
\hline
& \text{None} &  (1.0, 1.0) & (1.0, 2.7) & (1.0, 9.0) \\
Homophily & \text{Moderate} & (3.2, 1.0)  & (3.2, 2.0) & (3.2, 4.0) \\
& \text{High} & (5.1, 1.0) & (5.1, 1.7) &  (5.1, 3.3) \\
\hline
\end{tabular}
\label{tab:sc}
\end{table}

To calibrate MDR, target values $\phi^{\text{MDR}} \in \{1.0,2.0,4.0\}$ were first selected under moderate homophily ($\tau=3.2$), and parameter vectors $\boldsymbol{\beta}$ were chosen to produce these levels. These $\boldsymbol{\beta}$ vectors were then held fixed across the other homophily settings ($\tau=1.0$ and $\tau=5.1$), which implies that the resulting $\phi^{\text{MDR}}$ values vary with $\tau$. Table~\ref{tb:params} reports the scenario-specific values of $\boldsymbol{\eta}$ and $\boldsymbol{\beta}$.

\begin{table}[ht!]
\centering
\caption{
$\boldsymbol{\eta}$ parameters used to generate the three levels of homophily 
($\tau \in \{\text{none}, \text{moderate}, \text{high}\}$), and 
$\boldsymbol{\beta}$ parameters used to generate the three levels of MDR 
($\phi^{\text{MDR}} \in \{\text{none}, \text{moderate}, \text{high}\}$). 
The simulation study considered all nine possible combinations.
}\label{tb:params}
\begin{tabular}{lcc}
\toprule
Level & $\boldsymbol{\eta} = (\eta_{1},\eta_{2})$ & $\boldsymbol{\beta} = (\beta_{1},\beta_{2},\beta_{3},\beta_{4}) $ \\
\midrule
None     & $(-4.41,\,\textcolor{white}{-}0.00)$ & $(0.000,\,0.000,\,0.000,\,\textcolor{white}{-}0.000)$ \\
Moderate & $(-3.60,\,-0.19)$ & $(0.126,\,0.064,\,0.010,\,-0.017)$ \\
High     & $(-3.27,\,-0.28)$ & $(0.230,\,0.031,\,0.018,\,-0.003)$ \\
\bottomrule
\end{tabular}
\end{table}

Having defined the network and MDR scenarios, we then drew RDS samples from each simulated network using the following design. Each RDS sample began with 7~seeds drawn from the stationary distribution described in Equation~\eqref{eq:statdist}. Recruitment then proceeded according to the transition probabilities in Equation~\eqref{eq:mcextended}, with each participant distributing 2~coupons without replacement until the sample grew to $n=200$ units. For each scenario, 15~networks were generated, and 80~RDS samples were drawn from each, yielding 1200~samples per scenario.

\subsection{Results}\label{sec:simresults}

This section presents the results from the nine simulated scenarios summarized in Table~\ref{tab:sc}. The primary objective is to compare the performance of existing estimators, $\hat{\mu}_{VH}^{II}$, $\hat{\mu}_{DR}^{II}$, $\hat{\mu}_{Lu}^{ego}$, and $\hat{\mu}_{DR}^{ego}$, against our extended estimators that incorporate Multiple Differential Recruitment (MDR), namely $\hat{\mu}_{MDR}^{II}$ and $\hat{\mu}_{MDR}^{ego}$. The Differential Recruitment (DR) estimators are calculated assuming that DR is driven solely by the outcome variable $\boldsymbol{z}$ since, by construction, they can only consider one categorical variable at a time. Comparisons are based on the point estimation error and the coverage rates of the 95\% confidence intervals.

\begin{figure}[ht!] 
    \centering
    \includegraphics[width=.95\textwidth]{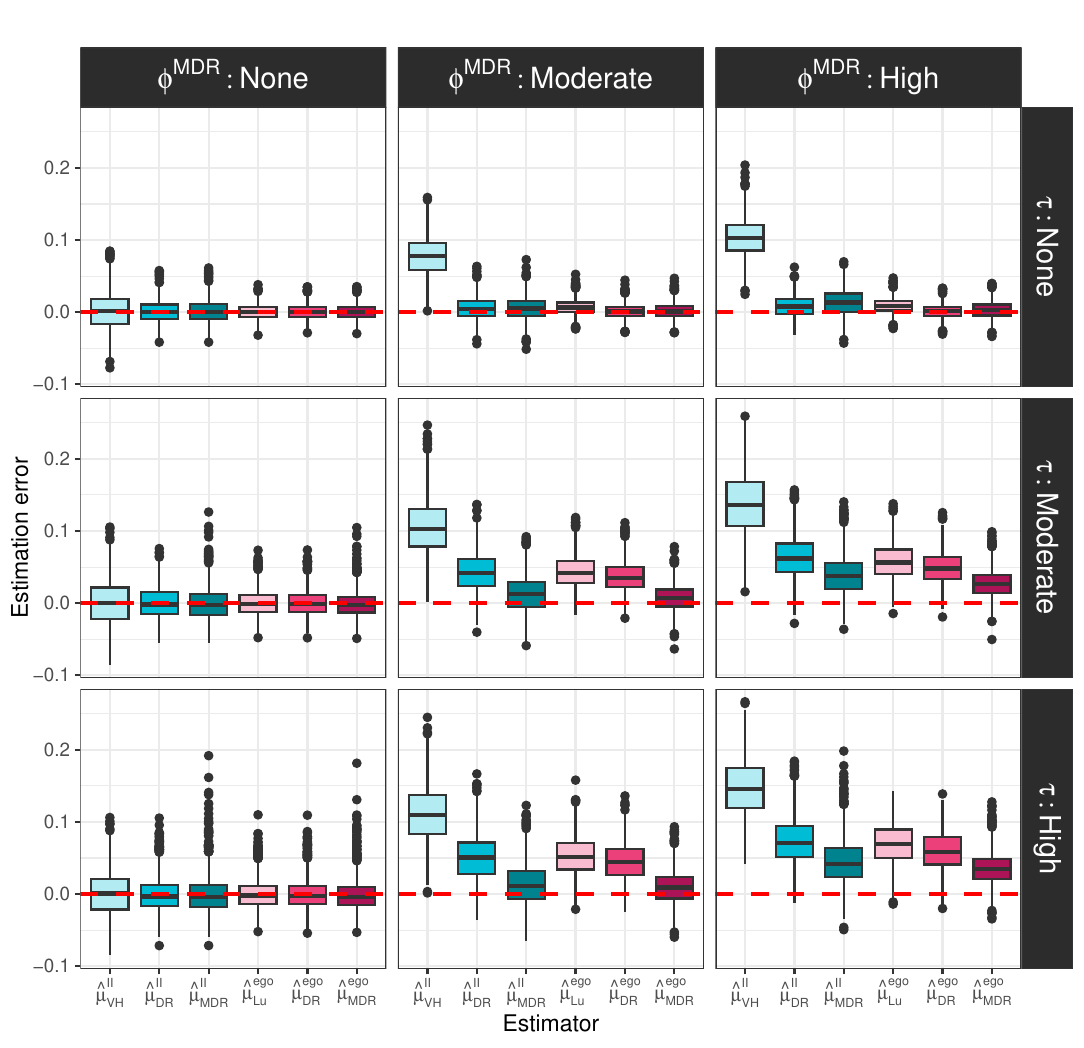}
\caption{Estimation error for each estimator in different scenario configurations. Rows correspond to homophily levels of homophily $\tau \in \{\text{none, moderate, high}\}$, and columns correspond to MDR levels, $\phi^{\text{MDR}} \in \{\text{none, moderate, high}\}$. Estimators are ordered from left to right as follows: $\hat{\mu}_{VH}^{II}, \hat{\mu}_{DR}^{II}, \hat{\mu}_{MDR}^{II}, \hat{\mu}_{Lu}^{ego}, \hat{\mu}_{DR}^{ego}$ and $\hat{\mu}_{MDR}^{ego}$. The dashed red horizontal line indicates zero estimation error.}
    \label{fig:sesgos}
\end{figure}

Figure~\ref{fig:sesgos} presents boxplots of the estimation errors of all six estimators across the nine scenarios. The rows correspond to the three levels of homophily, $\tau \in \{\text{none}, \text{moderate}, \text{high}\}$, while the columns represent the three levels of MDR, $\phi^{\text{MDR}} \in \{\text{none}, \text{moderate}, \text{high}\}$. The horizontal dashed line at zero represents the absence of error. In the top-left panel, representing the scenario without homophily or MDR, all estimators are approximately unbiased. However, the $II$-type estimators exhibit greater variability than the ego-type estimators, with $\hat{\mu}_{VH}^{II}$ being the most variable. 

As the level of MDR increases (moving along the first row), the bias of $\hat{\mu}_{VH}^{II}$ rises sharply, reflecting the fact that this estimator assumes random recruitment rather than differential or multiple differential recruitment. Consistent with the findings of \cite{Beaudry2020}, the VH estimator is highly sensitive to non-random recruitment. In contrast, the Lu estimator is far less affected by violations of the random recruitment assumption. The extended VH estimators, $\hat{\mu}_{DR}^{II}$ and $\hat{\mu}_{MDR}^{II}$, also show increasing bias with MDR, though to a much lesser extent. In contrast, the ego-type estimators $\hat{\mu}_{DR}^{ego}$ and $\hat{\mu}_{MDR}^{ego}$ remain nearly unbiased and display low variability across all three panels of the first row.

In the absence of MDR (first column), all estimators remain approximately unbiased across both moderate and high homophily levels ($\tau = 3.2$ and $5.1$, respectively). However, all estimators exhibit a slight increase in variability due to network homophily, regardless of its level. Notably, the $\hat{\mu}_{VH}^{II}$ estimator shows greater variability, suggesting it is more sensitive to network homophily.

The introduction of MDR in homophilous networks (moving across the second and third rows) biases all estimators and slightly increases their variance. This rise in bias across all estimators is attributable to the combined effects of homophily and MDR. However, the newly proposed MDR-adjusted estimators, $\hat{\mu}_{MDR}^{II}$ and $\hat{\mu}_{MDR}^{ego}$, consistently yield the smallest bias because they explicitly account for MDR in their formulations. In particular, under moderate MDR, these estimators remain nearly unbiased even in the presence of high homophily. Again, their modest increase in variability is due to the additional uncertainty introduced by estimating $\boldsymbol{\beta}$.

The Root Mean Square Error (RMSE) for all estimators across the nine scenarios is reported in Table~\ref{tab:RMSE}. These results complement the analysis presented in Figure~\ref{fig:sesgos}. In each scenario, the estimator with the smallest RMSE is indicated in bold and marked with an asterisk ($^\ast$). A Bonferroni test was also applied to determine whether the estimator with the lowest RMSE differed significantly from the others at a 95\% familywise confidence level. Estimators for which no significant difference was detected relative to the minimum RMSE value are indicated in bold without an asterisk.

\begin{table}[ht!]
\centering
\caption{Root-Mean-Square Error (RMSE) across estimators and scenarios. The values in bold accompanied by an asterisk ($^\ast$) correspond to the estimators attaining the lowest RMSE. Furthermore, values in bold indicate estimators for which the Bonferroni-adjusted comparison provides no statistically significant evidence that their mean squared error differs from that of the estimator with the lowest RMSE at a 95\% familywise confidence level.}
\begin{tabular}{lcccccc}
\toprule 
\multirow{2}{*}{Scenario : $(\tau,\phi^{\text{MDR}})$} & \multicolumn{6}{c}{Estimator} \\
\cmidrule{2-7}
& $\hat{\mu}_{VH}^{II}$ & $\hat{\mu}_{DR}^{II}$ & $\hat{\mu}_{MDR}^{II}$ & $\hat{\mu}_{Lu}^{ego}$ & $\hat{\mu}_{DR}^{ego}$ & $\hat{\mu}_{MDR}^{ego}$ \\ 
\midrule
1 : (None, None) & 0.0258 & 0.0151 & 0.0153 & \textbf{0.0106} & \textbf{0.0105*} & \textbf{0.0106}  \\
2 : (None, Moderate) & 0.0825 & \justifying 0.0158 & 0.0172 & 0.0126 & \textbf{0.0102*} & 0.0114   \\
3 : (None, High) & 0.1068 &  0.0169 & 0.0224 & 0.0132 & \textbf{0.0096*} & 0.0119   \\
4 : (Moderate, None) & 0.0306 & 0.0219 & 0.0231 & \textbf{0.0183}  & \textbf{0.0181*} & \textbf{0.0185} \\
5 : (Moderate, Moderate) & 0.1119 & 0.0520 & 0.0283 & 0.0492 & 0.0426 & \textbf{0.0196*} \\
6 : (Moderate, High) & 0.1446 & 0.0708 & 0.0481 & 0.0629 & 0.0542 & \textbf{0.0341*} \\
7 : (High, None) & 0.0311 & 0.0248 & 0.0274 & \textbf{0.0214*} & \textbf{0.0216} & \textbf{0.0232}  \\
8 : (High, Moderate) & 0.1176 & 0.0601 & 0.0327 & 0.0589 & 0.0519 & \textbf{0.0248*} \\
9 : (High, High) & 0.1526 & 0.0813 & 0.0562 & 0.0755 & 0.0658 & \textbf{0.0433*} \\
\bottomrule
\end{tabular}
\label{tab:RMSE}
\end{table}

In the absence of MDR (scenarios 1, 4, and 7), the ego-type estimators ($\hat{\mu}_{Lu}^{ego}, \hat{\mu}_{DR}^{ego},$ and $\hat{\mu}_{MDR}^{ego}$) achieved the best performance, regardless of the homophily levels ($\tau \in \{\text{none, moderate, high}\}$). These results are consistent with \cite{Beaudry2020}, which found that the Lu estimator ($\hat{\mu}_{Lu}^{ego}$) and its DR extension ($\hat{\mu}_{DR}^{ego}$) exhibit less variability in the presence of homophily than the $II$-type estimators. Our findings further show that the MDR extension, $\hat{\mu}_{MDR}^{ego}$, retains this property and performs equally well under these conditions.

In the presence of MDR ($\phi^{\text{MDR}} \in \{\text{moderate, high}\}$) but in the absence of homophily (scenarios 2 and 3), $\hat{\mu}_{DR}^{ego}$ achieves the best performance, with $\hat{\mu}_{MDR}^{ego}$ closely following as the second-best estimator in terms of RMSE. Although the Bonferroni test indicates that the difference between these two estimators is statistically significant in both cases, the magnitude of the difference may be negligible for practical purposes.

Finally, $\hat{\mu}_{MDR}^{ego}$ performs best in scenarios 5, 6, 8, and 9, which feature both moderate to high MDR ($\phi^{\text{MDR}}~\in~\{\text{moderate, high}\}$) and homophily ($\tau$~$\in$~$\{\text{moderate, high}\}$). The Bonferroni test suggests a statistically significant difference in the mean squared error of $\hat{\mu}_{MDR}^{ego}$ and $\hat{\mu}_{MDR}^{II}$. Overall, $\hat{\mu}_{MDR}^{ego}$ achieves the best performance under those conditions.

\begin{figure}[ht!] 
    \centering
    \includegraphics[width=.95\textwidth]{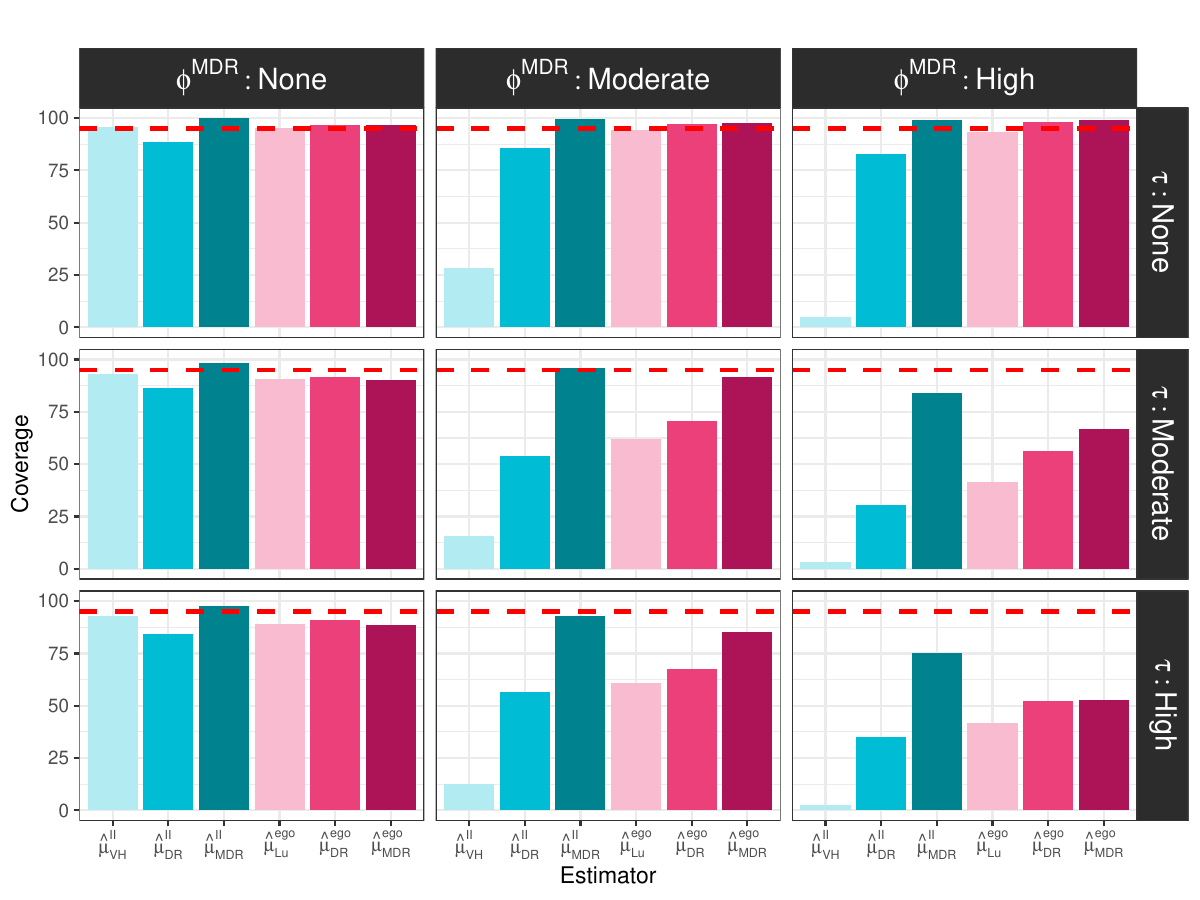}
    \caption{95\% confidence interval coverage by estimator in different scenarios configuration. Rows represents the three different levels of homophily $\tau \in \{\text{none, moderate, high}\}$, and the columns the MDR magnitude $\phi^{\text{MDR}} \in \{\text{none, moderate, high}\}$. The order of the estimators from left to right are: $\hat{\mu}_{VH}^{II},\hat{\mu}_{DR}^{II},\hat{\mu}_{MDR}^{II},\hat{\mu}_{Lu}^{ego},\hat{\mu}_{DR}^{ego}$ and $\hat{\mu}_{MDR}^{ego}$. The dashed horizontal red line represents the nominal coverage level of 95\%.}
    \label{fig:cob}
\end{figure}

Figure~\ref{fig:cob} reports the 95\% confidence interval coverage of all prevalence estimators. The layout mirrors that of Figure~\ref{fig:sesgos}: homophily levels vary across rows, MDR magnitudes across columns, and estimators are displayed in the same order. The dashed red horizontal line indicates the nominal 95\% coverage level. The confidence intervals are constructed using the variance estimation methods described in Sections~\ref{sec:bootlit} and~\ref{sec:bootnb}. 

In particular, the existing estimators ($\hat{\mu}_{VH}^{II}$, $\hat{\mu}_{DR}^{II}$, $\hat{\mu}_{Lu}^{ego}$, $\hat{\mu}_{DR}^{ego}$) rely on the Salganik bootstrap procedure or its extensions, whereas the MDR-adjusted estimators ($\hat{\mu}_{MDR}^{II}$, $\hat{\mu}_{MDR}^{ego}$) uses the modified NB procedure (see Algorithm \ref{alg1}). To ensure comparability, all margins of error in the confidence intervals are based on the standard normal quantiles ($z_{1-\alpha/2}$), regardless of the original methodology

In the top-left panel, when there is no homophily in age or MDR, the coverage of all variance estimation methods reaches the nominal 95\% confidence level, except for $\hat{\mu}_{DR}^{II}$, which has approximately 88\% coverage. This undercoverage likely arises because the DR is induced directly by the outcome variable $\boldsymbol{z}$, rather than by an auxiliary covariate $u$ associated with $\boldsymbol{z}$. As shown by \cite{Beaudry2020}, when the response variable coincides with the variable used to induce the DR, the performance of the extended VH estimator $\hat{\mu}_{DR}^{II}$ deteriorates modestly.

We also observe that the coverage for $\hat{\mu}_{MDR}^{II}$ reaches 100\% in the scenario without MDR or homophily. This conservative result stems from the modified NB overestimating the estimator's variance, specifically the variability of the $\beta_2$ parameter. This is of particular interest as $\beta_2$ is associated with infection status and affects the stationary distribution of infected versus uninfected nodes differently. This discrepancy produces greater variability in the resampled point estimates of $\hat{\mu}_{MDR}^{II}$ and, consequently, yields wider confidence intervals than expected. 

This overestimation of the variance is not observed for $\hat{\mu}_{MDR}^{ego}$. As shown in the Result~\ref{res:LuyVH} of the Appendix~\ref{appendix}, $\hat{\mu}_{MDR}^{ego}$ is related to $\hat{\mu}_{MDR}^{II}$ through a scaling constant $c$ and the first-order Taylor expansion around $\mu$ reveals that $\operatorname{Var}(\hat{\mu}_{MDR}^{ego}) \approx f(c)\operatorname{Var}(\hat{\mu}_{MDR}^{II})$. As the variability of $\hat{\beta}_2$ increases, $c$ often exceeds 1, which causes $f(c)$ to fall between 0 and 1. Consequently, $\hat{\mu}_{MDR}^{ego}$ yields a smaller estimated variance and, by extension, a lower coverage rate than $\hat{\mu}_{MDR}^{II}$.

Introducing moderate or high MDR (first-row) causes a decline in the 95\% confidence interval coverage for both $\hat{\mu}_{VH}^{II}$ and $\hat{\mu}_{DR}^{II}$. This loss of coverage is consistent with the increase in bias observed in Figure~\ref{fig:sesgos}. The decline is more prominent for $\hat{\mu}_{VH}^{II}$, whose bias grows substantially under MDR, while the reduction in coverage for $\hat{\mu}_{DR}^{II}$ is relatively mild. As expected, the increase in estimator bias directly translates into poorer interval coverage.

In the scenario where homophily is moderate or high ($\tau\in\{3.2,5.1\}$) and MDR is absent (the second and third rows of the first column), all methods achieve reasonable coverage. However, as MDR increases along these rows, coverage decreases for all estimators. This decline is directly linked to the growing bias shown in Figure~\ref{fig:sesgos}, as the interaction between homophily and MDR significantly exacerbates bias in estimators that do not adjust for MDR. Despite this overall decline, our MDR-based estimators consistently achieve the highest coverage. The modified NB slightly underestimates the variance of $\hat{\mu}_{MDR}^{ego}$, which explains why its coverage remains lower than that of $\hat{\mu}_{MDR}^{II}$.

In conclusion, $\hat{\mu}_{MDR}^{ego}$ shows the best overall performance in terms of RMSE, outperforming alternative estimators in all but two scenarios. Both $\hat{\mu}_{MDR}^{II}$ and $\hat{\mu}_{MDR}^{ego}$ consistently yield coverage rates that are either better than or equivalent to existing methodologies. However, $\hat{\mu}_{MDR}^{II}$ tends to produce conservative confidence intervals in the absence of either homophily or MDR. Conversely, $\hat{\mu}_{MDR}^{ego}$ rarely yields conservative intervals but exhibits lower coverage than $\hat{\mu}_{MDR}^{II}$ when both homophily and MDR are present. Overall, our MDR-based estimators show a clear advantage in both precision and reliability. These results suggest that collecting data to measure MDR can significantly improve inference from RDS data.

\section{Application}\label{sec:5}

We apply our Multiple Differential Recruitment (MDR) methodology to an RDS study that was conducted in Chile in 2022 \citep{PAP2024}. The objective of that study was to collect information on Venezuelan immigrants' political participation in Chilean politics and the factors motivating their engagement. The target population consisted of Venezuelan nationals aged 18 or older residing in the Metropolitan Region of Santiago, Chile. 

The RDS sample contained 32 seeds, selected by four organizations with extensive experience working with Venezuelan communities in Chile. Each seed and subsequent respondents could recruit up to three of their contacts. Due to a few administrative errors, some participants were recruited beyond this limit. The final cleaned dataset consisted of 375 participants.

\begin{figure}[ht!] 
    \centering
    \includegraphics[width=0.65\textwidth]{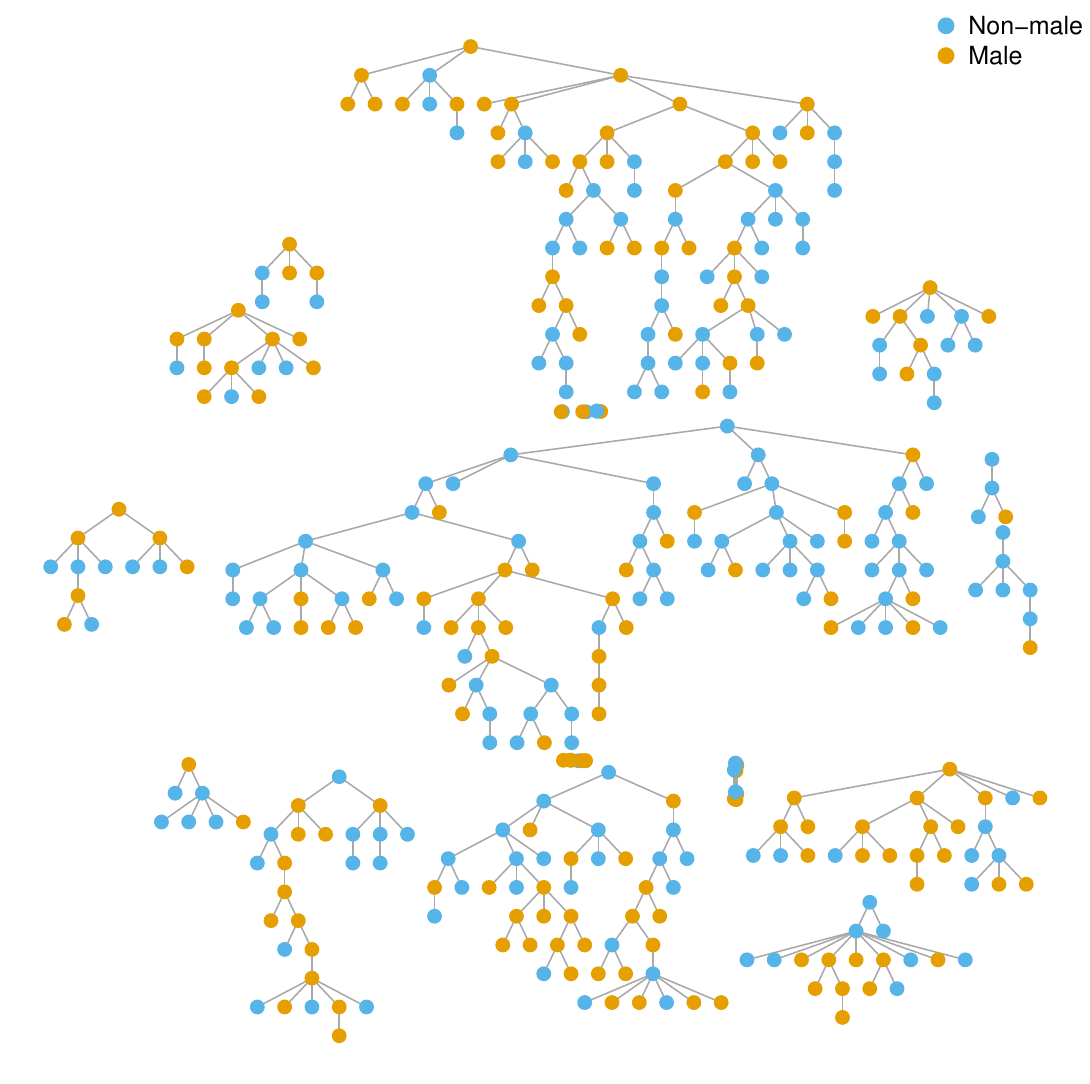}
\caption{Tree of the RDS sampling process, the non-males nodes are colored with dark gray and the males with light gray.}
    \label{fig:rdstree}
\end{figure}

The survey included common RDS questions, such as participants' degree (i.e., the number of people from the target population with whom they had contact in the previous week), gender, and year of birth. We approximated the respondents' age at the time of the survey based on the latter. In addition to those characteristics, the study collected information on participants' social networks within the target population. Specifically, respondents reported the number of their contacts by gender (degree by gender) and age group (degree by age group), with age categories defined in 5-year intervals, except for the youngest (18–19 years) and oldest (over 80 years) groups. Figure~\ref{fig:rdstree} shows the RDS sampling tree, with non-male nodes colored dark gray and male nodes colored light gray.

We performed simple imputation for the single missing gender value based on observed recruitment patterns. In particular, we found that respondents recruited same-gender contacts with probability 0.62 in the sample, so we imputed the missing gender accordingly. In the final sample, 49\% were male, and 51\% were non-male. 

We identified inconsistencies between observed recruitment and reported degrees (overall degree, degree by gender, and degree by age group) and resolved them as follows. First, we ensured that each participant's reported degree was at least as large as their observed recruitment activity, counting both their recruiter and any recruits. For example, a participant who was recruited and subsequently recruited two others was assigned degree three, even if they reported zero contacts.

Second, when discrepancies occurred across the three degree measures, we treated the most frequent value as the true total degree and adjusted the remaining degrees accordingly. For example, if the overall and gender-specific degrees were both 4 but the age-specific degree was 3, we increased the age-specific degree to 4. Age-specific degrees were corrected by adding or removing contacts based on probabilities that favored age categories closer to the participant's age. Specifically, the probability of adding a contact to the age group $g \in \left\{ [18, 20), [20, 25), \dots, [80, \infty) \right\}$ was calculated as $pr_g \propto (d(g_i,g) \cdot f_g)^{-1}$, where $d(g_i,g)$ represents the distance in years from the participant's group $g_i$ (with $d(g_i, g_i) = 1$ and $d([40, 45), [30, 35)) = 10$, for instance). Consequently, the model favors the participant's own age bracket, subject to the group-frequency weights $f_g$. For contact removal, the selection probabilities were proportional to $pr_g^{-1}$, restricted to categories with at least one reported contact. 

Gender-specific degrees were corrected based on observed recruitment patterns, assigning probability 0.62 to same-gender contacts when adding a contact and to opposite-gender contacts when removing a contact. In total, 48 of the 375 cases were modified to ensure internal consistency.

Implementing the MDR methodology requires ages and genders for participants' contacts, but the data provide only aggregate counts by age category and gender. We therefore simulated individual contact ages by drawing uniformly within each reported age interval (e.g., $X \sim \mathrm{Uniform}(20,25)$ for a contact in the 20--24 age group). Gender labels were then randomly assigned to each contact, ensuring that the total gender allocation matched the reported gender distribution. Using this completed dataset, we computed prevalence and variance estimators under the random recruitment, DR, and MDR methodology. This completed dataset is approximate, and the analysis should be interpreted as an illustration of the MDR framework under partial alter information, rather than the ideal case of fully observed covariates.

For the DR estimators, gender was assumed to be the sole recruitment-related covariate, coded as a binary indicator (1 for male, 0 otherwise). The estimated DR magnitude was 1.19, indicating that males were recruited with approximately 19\% higher probability than non-males. In the MDR model, recruitment was modeled as a function of (i) the potential recruit's age, (ii) the potential recruit's gender, and (iii) the absolute age difference between the recruiter and the potential recruit. 

The maximum-likelihood estimates for the coefficients were 0.028, 0.19, and 0.012, respectively. To illustrate the interpretation of these coefficients, consider a recruiter $i$ aged 35 who faces two potential recruits: $j_1$, a 40-year-old male, and $j_2$, a 30-year-old female. According to Equation~\eqref{eq:mcextended}, the relative probability of recruiting $j_1$ rather than $j_2$ is given by:
\begin{equation}
\frac{P_{ij_1}^{\text{MDR}}}{P_{ij_2}^{\text{MDR}}}
=
\frac{\exp(0.028\cdot 40 + 0.19\cdot 1 + 0.012\cdot 5)}
{\exp(0.028\cdot 30 + 0.19\cdot 0 + 0.012\cdot 5)}
\approx 1.6.
\end{equation}

Thus, the fitted MDR model implies a higher probability of recruiting an older male relative to a younger non-male, holding the absolute age difference constant. Also, the MDR gender coefficient is largely consistent with the DR effect; that is, assuming everything else is constant, there is approximately a 21\% higher chance of recruiting a male under MDR than a non-male. Having specified the recruitment models, we next estimated the proportion of males in the target population using each estimator. Table~\ref{tab:App} reports the resulting point estimates, standard errors, and 95\% confidence intervals.

\begin{table}[ht!]
\centering
\caption{Point estimates of the proportion of males in the Venezuelan population residing in the Metropolitan Region of Santiago, along with the standard errors and 95\% confidence intervals (CI) across estimators. Results are based on the original dataset.}
\begin{tabular}{lcccccc}
\cmidrule{2-7}
 & $\hat{\mu}_{VH}^{II}$ & $\hat{\mu}_{DR}^{II}$ & $\hat{\mu}_{MDR}^{II}$ & $\hat{\mu}_{Lu}^{ego}$ & $\hat{\mu}_{DR}^{ego}$ & $\hat{\mu}_{MDR}^{ego}$ \\
\midrule
Prevalence estimate  & 0.489 & 0.430 & 0.437 & 0.456 & 0.439 & 0.452  \\
Std. error           & 0.042 & 0.041 & 0.045 & 0.035 & 0.041 & 0.035  \\
Lower bound CI             & 0.406 & 0.349 & 0.350 & 0.389 & 0.358 & 0.383  \\
Upper bound CI  & 0.572 & 0.511 & 0.525 & 0.524 &  0.519 & 0.522 \\
\bottomrule
\end{tabular}
\label{tab:App}
\end{table}

Given the relatively low magnitude of the MDR coefficients (Table~\ref{tb:CIBapp}), we expect the point estimates and standard errors to be broadly similar across methods, which is consistent with the results in Table~\ref{tab:App}. Among the estimators considered, $\hat{\mu}_{VH}^{II}$ is the most sensitive to recruitment bias in our simulation study and also yields the largest point estimate in this application. Finally, the estimators with the smallest standard error, $\hat{\mu}_{Lu}^{ego}$ and $\hat{\mu}_{MDR}^{ego}$, are both of the ego-type, which aligns with the results in Section~\ref{sec:simresults}.

\begin{table}[ht!]
\centering
\caption{Point estimates and 95\% confidence intervals (CI) for the MDR parameters. Results are derived using the original dataset and the modified NB method.}\label{tb:CIBapp}
\begin{tabular}{lccc}
\toprule
Covariate & Estimate & Lower bound CI & Upper bound CI \\
\midrule
Age      &0.028 & 0.010 & 0.046  \\
Gender     &0.190 & -0.090 & 0.468 \\
Age difference & 0.012 & -0.013 &  0.037 \\
\bottomrule
\end{tabular}
\label{tab:Ci1}
\end{table}

The 95\% confidence intervals in Table~\ref{tb:CIBapp} indicate that two of the three MDR parameters (gender and age difference) are not statistically significant at the $\alpha=0.05$ level. To further assess the impact of recruitment bias, we performed a sensitivity analysis using a scenario with stronger MDR effects. We created a modified dataset by perturbing the recruitment covariates so that all three MDR coefficients were statistically significant at the $\alpha=0.05$ level. Specifically, we adjusted both the alters' gender and age differences between recruiters and their alters.

First, to induce a stronger preference for recruiting males, we modified the reported gender composition of recruiters' contacts by converting 70\% of their non-recruited alters to non-male alters. For example, suppose a participant reports 12 contacts, of whom 2 were recruited into the sample, and the remaining 10 are split evenly between males and non-males. Under our modification, these 10 unrecruited contacts are reassigned to 3 males and 7 non-males. Recruiting a male implies a stronger deviation from random recruitment, since males now constitute a smaller share of the alters.

Second, as with gender, we sought to strengthen the recruitment effect related to age differences. Because the original coefficient was positive (0.012), indicating a tendency to recruit contacts with larger age gaps, we reduced each contact's age difference by 3 years to accentuate this effect. For example, if a 20-year-old had a 27-year-old contact, the contact’s age would be adjusted to 24. Under this modified composition, recruiting an older contact (e.g., age 35) reflects a stronger age-difference effect, since most contacts are now closer in age to the recruiter.  These modifications resulted in all MDR parameters now being statistically significant at a $\alpha = 0.05$ level, as shown in Table~\ref{tb:CIBappUpdated}. 

\begin{table}[ht!]
\centering
\caption{Point estimates and 95\% confidence intervals (CI) for the MDR parameters. Results are derived using the modified dataset and the modified NB method.}\label{tb:CIBappUpdated}
\begin{tabular}{lccc}
\toprule
Covariate &Estimate & Lower bound CI & Upper bound CI \\
\midrule
Age      &$0.044$ & 0.018 & 0.070   \\
Gender & $0.655$ & 0.429 & 0.881  \\
Age difference   & $0.076$ & 0.044 & 0.109 \\
\bottomrule
\end{tabular}
\label{tab:ci2}
\end{table}

Table~\ref{tab:Mod} reports the estimated proportion of males under the modified dataset. We note that the results for the $\hat{\mu}_{VH}^{II}$ methodology are identical to those of the original dataset (Table~\ref{tab:App}) since this estimator solely depends on the individuals' degrees and their outcome variable, neither of which were modified.

\begin{table}[ht!]
\centering
\caption{Point estimate of the proportion of males in the Venezuelan population residing in the Metropolitan Region of Santiago together with the standard error and the 95\% confidence intervals across estimators. Results are based on the modified dataset.}
\begin{tabular}{lcccccc}
\cmidrule{2-7} 
 & $\hat{\mu}_{VH}^{II}$ & $\hat{\mu}_{DR}^{II}$ & $\hat{\mu}_{MDR}^{II}$ & $\hat{\mu}_{Lu}^{ego}$ & $\hat{\mu}_{DR}^{ego}$ & $\hat{\mu}_{MDR}^{ego}$ \\
\midrule
Prevalence estimate  & 0.489 & 0.292  & 0.302 & 0.404 & 0.341 & 0.365 \\
Std. error     & 0.042 & 0.039  &  0.036 & 0.035 & 0.039 & 0.034   \\
Lower bound CI &  0.406 & 0.216  & 0.230 & 0.334 & 0.265 & 0.298  \\
Upper bound CI & 0.572 & 0.368 & 0.373 & 0.473 &  0.416 & 0.433 \\
\bottomrule
\end{tabular}
\label{tab:Mod}
\end{table}

In contrast, the remaining five estimators are affected by the significant recruitment bias. In particular, we expect prevalence estimates to decrease because the modified alter information makes male participants appear disproportionately likely to have been sampled compared to the original dataset. For the DR and MDR estimates, this artificially inflates the estimated sampling probabilities for males and, as a result, decreases their weight in the estimators. For $\hat{\mu}_{Lu}^{ego}$, the decrease is attributable to the structure of the estimator rather than a direct update to the sampling weights. These values should be interpreted strictly as a measure of the estimators' sensitivity to large MDR, rather than as plausible population figures.

We also note that the decrease in the prevalence estimate after the introduction of stronger MDR is larger for $\hat{\mu}_{DR}^{II}$ and $\hat{\mu}_{MDR}^{II}$ than the corresponding ego-estimators. This result is attributable to the large MDR effect on the gender variable, which coincides with the outcome variable, and to the structure of the estimator (Hájek-type estimators), which makes these estimators more sensitive to changes in sampling weights related to the outcome variable. We also note a reduction in standard error for most estimators. This reduction is likely a numerical consequence of the lower estimated prevalence following the MDR adjustment. Finally, the lowest standard error is achieved for $\hat{\mu}_{MDR}^{ego}$, consistent with our simulation study in the presence of moderate to high MDR.

In conclusion, we applied our methodology to a dataset of Venezuelans aged 18 and older residing in the Metropolitan Region of Santiago, Chile, using gender as the response variable to estimate the proportion of males in 2022. With the original dataset, differences across estimators were minimal. However, introducing stronger MDR effects on gender, age, and age differences led to noticeable changes in point estimates, with standard errors also affected but to a lesser extent. These results highlight that collecting additional information on participants’ alters can help identify recruitment biases, evaluate their potential impact, and may lead to more reliable prevalence estimates.

\section{Discussion} \label{sec:6}

Respondent-Driven Sampling (RDS) is a method designed to address data collection challenges in hard-to-reach populations. Because these populations typically lack a sampling frame, the sampling is conducted through peer recruitment. As a result, the sample is constrained by the network's structure and governed by respondents' recruitment behaviors.

In this work, we incorporate variables that may influence participants' recruitment behaviors into the inferential framework, including the characteristics of potential recruits and features of the relationship between recruiters and potential recruits. By doing so, the MDR methodology provides a novel and more realistic representation of the sampling process, moving beyond both the oversimplistic assumption of random recruitment and the limitations of univariate differential recruitment. Ultimately, this approach helps reduce non-sampling errors in RDS statistical inference.

Early RDS representations are based on a Markov chain (MC) over the state space of nodes in the target population, where the chain's transition matrix models participants' recruitment. Under this framework, all potential recruits of a given recruiter are assumed to have an equal probability of being selected. Hence, recruitment depends solely on the recruiter's degree, rather than exogenous covariates. Differential recruitment (DR) relaxes this assumption by allowing unequal transition probabilities. However, the DR framework remains limited, as its MC can accommodate only a single categorical variable.

To generalize this framework, we introduce an MDR approach that allows for a more flexible parameterization of recruitment probabilities. Specifically, we define an MC whose transition matrix depends on the characteristics of potential recruits and the relationship between the recruiter and those recruits. In this model, covariates are weighted by unknown parameters determined through maximum likelihood estimation (MLE).

To obtain the MDR prevalence estimators, we derived the MC stationary distribution. This distribution depends on the MDR parameters. By replacing the unknown parameters with their corresponding MLE estimates, we obtained the estimated stationary distribution (up to a constant of proportionality). This distribution serves as the sampling probabilities used to construct our MDR prevalence estimators, $\hat{\mu}_{MDR}^{II}$ and $\hat{\mu}_{MDR}^{ego}$, and as such, they explicitly account for multivariate recruitment biases. To measure the uncertainty of the proposed estimators, we slightly modified the original neighborhood bootstrap (NB) procedure by fixing the replicate size to match the original RDS sample size.

We conduct a simulation study to evaluate the performance of the proposed methodology against estimators developed under random and DR schemes. In the simulations, we generate populations exhibiting varying levels of age-based homophily. In addition, we consider different sampling regimes designed to capture a range of MDR magnitudes.

Our results indicate that in the absence of MDR, all estimators are approximately unbiased. However, the ego-type estimators ($\hat{\mu}_{Lu}^{ego}$, $\hat{\mu}_{DR}^{ego}$, and $\hat{\mu}_{MDR}^{ego}$) exhibit lower variability than the II-type estimators ($\hat{\mu}_{VH}^{II}$, $\hat{\mu}_{DR}^{II}$, and $\hat{\mu}_{MDR}^{II}$), a finding that holds regardless of the degree of homophily. Furthermore, in the absence of homophily, most estimators remain largely unbiased across all MDR levels. The main exception is $\hat{\mu}_{VH}^{II}$, which, as demonstrated in our simulations, is highly sensitive to MDR. Estimators of the ego-type remain more efficient under those conditions. 

When we introduce MDR in homophilous networks, bias is significantly exacerbated for all methods. However, under these conditions, $\hat{\mu}_{MDR}^{II}$ and $\hat{\mu}_{MDR}^{ego}$ exhibit the lowest bias among all tested estimators. Overall, $\hat{\mu}_{MDR}^{ego}$ emerges as the estimator with the best RMSE performance. Aside from two scenarios where it is equivalent to the best estimator for practical purposes, it is consistently in the set of the best estimators.

The results for the coverage of the 95\% confidence intervals are largely consistent with the RMSE results. All estimators perform similarly without MDR, but the coverage for $\hat{\mu}_{VH}^{II}$ sharply declines when MDR is introduced. Furthermore, the coverage rates for our proposed estimators tend to be higher than those of alternative methods. The conservative coverage of $\hat{\mu}_{MDR}^{II}$ when either homophily or MDR is present (first row or first column) is mainly attributable to the overestimation of the variance by the modified NB. 

In contrast, the undercoverage of $\hat{\mu}_{MDR}^{ego}$ when both are present is due to mild bias in the prevalence estimation and to the modified NB underestimation of its variance. Therefore, the MDR estimators achieve either superior or equivalent coverage compared to other methods. However, $\hat{\mu}_{MDR}^{II}$ is preferable in the challenging scenarios where both MDR and homophily occur simultaneously.

We apply the proposed methods to an RDS study conducted among Venezuelan immigrants (aged 18 and older) residing in the Metropolitan Region of Santiago, Chile. Our inferential goal is to estimate the proportion of males in that population. The methodology is implemented on both the original dataset and a modified version designed to amplify the magnitude of the MDR.

In the original dataset, where MDR is low to moderate, point estimates across all methods are largely similar, consistent with our simulation results. However, the ego-type estimators, specifically $\hat{\mu}_{MDR}^{ego}$ and $\hat{\mu}_{Lu}^{ego}$, achieve the highest precision. In contrast, MDR is more pronounced in the modified dataset, corresponding to a moderate to high MDR scenario. The results show that, except for $\hat{\mu}_{VH}^{II}$, all point estimates decrease relative to those from the original dataset. This occurs because simulated oversampling of males increases their sampling probability, thereby reducing their individual weight in the final estimate. 

In addition, we also find that our estimator $\hat{\mu}_{MDR}^{ego}$ shows the lowest standard error, followed by $\hat{\mu}_{Lu}^{ego}$ and $\hat{\mu}_{MDR}^{II}$. This result is mostly consistent with the simulation study results for scenarios with moderate to high MDR. Overall, the empirical results from both datasets corroborate the findings from the simulation studies: estimators are sensitive to recruitment biases, and $\hat{\mu}_{MDR}^{ego}$ shows the smallest standard error in the presence of MDR.

In conclusion, we propose a novel RDS model that explicitly incorporates variables that may influence participants' recruitment patterns. This framework allows inference while accounting for factors that influence their recruitment decisions, rather than assuming random recruitment. The results from the simulation study and the empirical application are encouraging, especially when recruitment is not random. This highlights the potential value of collecting additional information on participants and their social networks in future RDS studies. 

The practical performance of the proposed approach depends on the availability of relevant covariate information on alters, or on the ability to approximate such information with sufficient accuracy. In practice, however, alter-level covariates in RDS studies are often partially observed or unavailable, requiring reconstruction or estimation. This process introduces additional uncertainty that is not explicitly accounted for in the current estimation framework and may affect the variability of the resulting estimators. Addressing incomplete alter-level information in MDR contexts, and developing methods to properly incorporate this uncertainty, remains an important direction for future research.

\section{Acknowledgments}
Vanesa Reinoso was funded by ANID-Subdirección de Capital Humano/Doctorado Nacional/2021-21211397, and Jonathan Acosta was partially supported by ANID/FONDECYT Initiation Grant No. 11230502. This work was also supported by the Agencia Nacional de Investigación y Desarrollo of Chile [ANID-Millennium Science Initiative Program-ICN17\_002]. We thank Instituto Milenio Fundamento de los Datos for providing the data used in the application of this work and Bruna Fonseca de Barros, Inés Fynn,  Lihuen Nocetto, Isabelle S. Beaudry,  Juan Pablo Luna, Rafael Piñeiro, and Fernando Rosenblatt Rodríguez for their roles in the data collection.

\bibliographystyle{apalike}
\bibliography{references}

@article{Hansen1943,
author = {Morris H. Hansen and William N. Hurwitz},
title = {On the theory of sampling from finite populations},
volume = {14},
journal = {The Annals of Mathematical Statistics},
number = {4},
pages = {333--362},
year = {1943},
doi = {10.1214/aoms/1177731356}
}

@article{Frank1986,
author = {Frank, Ove and Strauss, David},
year = {1986},
number = {395},
pages = {832--842},
title = {Markov graphs},
volume = {81},
journal = {Journal of the American Statistical Association},
doi = {10.2307/2289017}
}

@article{Heckathorn1997,
author = {Heckathorn, Douglas},
year = {1997},
pages = {174--199},
title = {Respondent-driven sampling: {A} new approach to the study of hidden populations},
volume = {44},
number = {2},
journal = {Social Problems},
doi = {10.2307/3096941}
}

@book{Degenne1999,
  title     = {Introducing social networks},
  author    = {Degenne, Alain and Fors{\'e}, Michel},
  year      = {1999},
  publisher = {Sage Publications},
  address   = {London}
}

@article{Gile2010,
author = {Krista J. Gile and Mark S. Handcock},
title ={Respondent-driven sampling: {A}n assessment of current methodology},
journal = {Sociological Methodology},
volume = {40},
number = {1},
pages = {285--327},
year = {2010},
doi = {10.1111/j.1467-9531.2010.01223.x}
}

@article{Liu2012,
author = {Liu, Hongjie and Li, Jianhua and Ha, Toan and Li, Jian},
year = {2012},
pages = {13--21},
title = {Assessment of random recruitment assumption in respondent-driven sampling in egocentric network data},
volume = {1},
number = {2},
journal = {Social Networking},
doi = {10.4236/sn.2012.12002}
}

@article{Shi2019,
author = {Shi, Yongren and Cameron, Christopher and Heckathorn, Douglas},
year = {2019},
number = {1},
pages = {3--33},
title = {Model-based and design-based inference: {R}educing bias due to differential recruitment in respondent-driven sampling},
volume = {48},
journal = {Sociological Methods \& Research},
doi = {10.1177/0049124116672682}
}

@article{Hunter2006,
title = {Inference in curved exponential family models for networks},
author = {Hunter, David R. and Handcock, Mark S.},
year = {2006},
doi = {10.1198/106186006X133069},
volume = {15},
number = {3},
pages = {565--583},
journal = {Journal of Computational and Graphical Statistics}
}

@article{Goel2009,
  author  = {Sharad Goel and Matthew J. Salganik},
  title   = {Respondent-driven sampling as {M}arkov chain {M}onte {C}arlo},
  journal = {Statistics in Medicine},
  year    = {2009},
  volume  = {28},
  number  = {17},
  pages   = {2202--2229},
  doi     = {10.1002/sim.3613},
  url     = {https://doi.org/10.1002/sim.3613}
}

@article{Gile2015,
  title = {Diagnostics for Respondent-driven Sampling},
  author = {Gile, Krista J. and Johnston, Lisa G. and Salganik, Matthew J.},
  journal = {Journal of the Royal Statistical Society: Series A (Statistics in Society)},
  year = {2015},
  volume = {178},
  number = {1},
  pages = {241–269},
  doi = {10.1111/rssa.12059},
  month = {January},
  pmid = {27226702},
  pmcid = {PMC4877136}
}

@article{Gile2018,
author = {Gile, Krista and Beaudry, Isabelle and Handcock, Mark and Ott, Miles},
year = {2018},
pages = {65--93},
title = {Methods for inference from respondent-driven sampling data},
volume = {5},
journal = {Annual Review of Statistics and Its Application},
doi = {10.1146/annurev-statistics-031017-100704}
}

@article{Salganik2006,
author = {Salganik, Matthew},
year = {2006},
number = {7},
pages = {98--112},
title = {Variance estimation, design effects, and sample size calculations for respondent-driven sampling},
volume = {83},
journal = {Journal of Urban Health},
doi = {10.1007/s11524-006-9106-x}
}

@article{McPherson2001,
 ISSN = {03600572, 15452115},
 URL = {http://www.jstor.org/stable/2678628},
 author = {Miller McPherson and Lynn Smith-Lovin and James M. Cook},
 journal = {Annual Review of Sociology},
 pages = {415--444},
 publisher = {Annual Reviews},
 title = {Birds of a feather: {H}omophily in social networks},
 urldate = {2025-07-07},
 volume = {27},
 year = {2001}
}

@article{Banze2024,
  author       = {Banze, {A\'}uria Ribeiro and Muleia, Rachid and Nuvunga, Samuel and Boothe, Makini and Baltazar, Cynthia Semá},
  title        = {Trends in {HIV} prevalence and risk factors among men who have sex with men in {M}ozambique: {I}mplications for targeted interventions and public health strategies},
  journal      = {BMC Public Health},
  year         = {2024},
  volume       = {24},
  number       = {1},
  pages        = {1185},
  doi          = {10.1186/s12889-024-18661-0},
  url          = {https://doi.org/10.1186/s12889-024-18661-0}
}

@book{Lusher2013,
  title     = {Exponential random graph models for social networks: {T}heory, methods and applications},
  editor    = {Lusher, Dean and Koskinen, Johan and Robins, Garry},
  year      = {2013},
  publisher = {Cambridge University Press},
  address   = {Cambridge, UK},
  isbn      = {9780521191096}
}

@article{Currarini2009,
 ISSN = {00129682, 14680262},
 URL = {http://www.jstor.org/stable/40263853},
 author = {Sergio Currarini and Matthew O. Jackson and Paolo Pin},
 journal = {Econometrica},
 number = {4},
 pages = {1003--1045},
 publisher = {[Wiley, The Econometric Society]},
 title = {An Economic Model of Friendship: {H}omophily, Minorities, and Segregation},
 urldate = {2025-08-11},
 volume = {77},
 year = {2009}
}

@article{Johnston2010,
  author    = {Lisa G. Johnston and Keith Sabin},
  title     = {Sampling hard-to-reach populations with respondent driven sampling},
  journal   = {Methodological Innovations Online},
  year      = {2010},
  volume    = {5},
  number    = {2},
  pages     = {38--48},
  doi       = {10.4256/mio.2010.0017}
}

@article{Salganik2004,
author = {Salganik, Matthew and Heckathorn, Douglas},
year = {2004},
number = {1},
pages = {193--240},
title = {Sampling and estimation in hidden populations using respondent-drive sampling},
volume = {34},
journal = {Sociological Methodology},
doi = {10.1111/j.0081-1750.2004.00152.x}
}

@article{Beaudry2020,
author = {Isabelle S. Beaudry and Krista J. Gile},
title = {{Correcting for differential recruitment in respondent-driven sampling data using ego-network information}},
volume = {14},
journal = {Electronic Journal of Statistics},
number = {2},
pages = {2678--2713},
year = {2020},
doi = {10.1214/20-EJS1718}
}

@article{Tomas2011,
author = {Amber Tomas and Krista J. Gile},
title = {{The effect of differential recruitment, non-response and non-recruitment on estimators for respondent-driven sampling}},
volume = {5},
journal = {Electronic Journal of Statistics},
pages = {899--934},
year = {2011},
doi = {10.1214/11-EJS630}
}

@article{Volz2008,
author = {Volz, Erik and Heckathorn, Douglas},
year = {2008},
number = {1},
pages = {79--97},
title = {Probability based estimation theory for respondent driven sampling},
volume = {24},
journal = {Journal of Official Statistics}
}

@article{Lu2013,
author = {Lu, Xin},
year = {2013},
pages = {669--685},
title = {Linked ego networks: {I}mproving estimate reliability and validity with respondent-driven sampling},
volume = {35},
journal = {Social Networks},
doi = {10.1016/j.socnet.2013.10.001}
}

@article{Verdery2015,
  author       = {Verdery, Ashton M. and Merli, M. Giovanna and Moody, James and Smith, Jeffrey A. and Fisher, Jacob C.},
  title        = {Brief Report: {R}espondent-driven Sampling Estimators Under Real and Theoretical Recruitment Conditions of Female Sex Workers in {C}hina},
  journal      = {Epidemiology},
  year         = {2015},
  volume       = {26},
  number       = {5},
  pages        = {661--665},
  month        = sep,
  doi          = {10.1097/EDE.0000000000000335},
}

@article{Lansky2007,
  author    = {Amy Lansky and Lillian A. Abdul-Quader and Cyprian Wejnert and
               Donald R. Hall and Deborah M. Finlayson and Linda A. Garfein and
               Patrick S. Sullivan},
  title     = {Developing an {HIV} Behavioral Surveillance System for Injecting Drug Users: {T}he National {HIV} Behavioral Surveillance System},
  journal   = {Public Health Reports},
  volume    = {122},
  number    = {Suppl 1},
  pages     = {48--55},
  year      = {2007},
  publisher = {SAGE Publications},
  doi       = {10.1177/00333549071220S108},
  pmcid     = {PMC1804107},
  pmid      = {17354525}
}

@article{Yauck2022,
  author       = {Yauck, Mamadou and Moodie, Erica E. M. and Apelian, Herak and Fourmigue, Alain and Grace, Daniel and Hart, Trevor A. and Lambert, Gilles and Cox, Joseph},
  title        = {Neighborhood Bootstrap for Respondent‑Driven Sampling},
  journal      = {Journal of Survey Statistics and Methodology},
  volume       = {10},
  number       = {2},
  pages        = {419--438},
  year         = {2022},
  doi          = {10.1093/jssam/smab057},
}

@article{Johnston2008,
  author  = {Johnston, Lisa G. and Malekinejad, Mohsen and Kendall, Carl and Iuppa, Irene M. and Rutherford, George W.},
  title   = {Implementation Challenges to Using Respondent-Driven Sampling Methodology for {HIV} Biological and Behavioral Surveillance: {F}ield Experiences in International Settings},
  journal = {AIDS and Behavior},
  volume  = {12},
  number  = {4},
  pages   = {S131--S141},
  year    = {2008},
  doi     = {10.1007/s10461-008-9413-1}
}

@article{Magnani2005,
  author    = {Robert Magnani and Keith Sabin and Tobi Saidel and Douglas D. Heckathorn},
  title     = {Review of sampling hard-to-reach and hidden populations for {HIV} surveillance},
  journal   = {AIDS},
  year      = {2005},
  volume    = {19},
  number    = {Suppl. 2},
  pages     = {S67--S72},
  doi       = {10.1097/01.aids.0000172879.20628.e1}
}

@article{Heckathorn2002RDS_Extensions,
  author    = {Douglas D. Heckathorn and Salaam Semaan and Robert S. Broadhead and James J. Hughes},
  title     = {Extensions of Respondent-Driven Sampling: {A} New Approach to the Study of Injection Drug Users Aged 18-25},
  journal   = {AIDS and Behavior},
  year      = {2002},
  volume    = {6},
  number    = {1},
  pages     = {55--67},
  doi       = {10.1023/A:1014528612685}
}

@Manual{R-base,
    title = {R: {A} language and environment for statistical computing},
    author = {{R Core Team}},
    organization = {R Foundation for Statistical Computing},
    address = {Vienna, Austria},
    year = {2025},
    url = {https://www.R-project.org/},
  }

@article{Arayasirikul2015,
  title   = {A Qualitative Examination of Respondent-Driven Sampling ({RDS}): {P}eer Referral Challenges Among Young Transwomen in the {S}an {F}rancisco Bay Area},
  author  = {Arayasirikul, Sean and Cai, Xiang and Wilson, Erin C.},
  journal = {JMIR Public Health and Surveillance},
  year    = {2015},
  volume  = {1},
  number  = {2},
  pages   = {e9},
  doi     = {10.2196/publichealth.4573},
  url     = {http://publichealth.jmir.org/2015/2/e9/},
  pmid    = {27227143}
}

@article{Takahashi2025,
  title   = {Transnational Political Participation of Undocumented {M}exican Immigrants in the {US}: {R}espondent-Driven Sampling with the Hard-to-Reach Population},
  author  = {Takahashi, Yuriko and Song, Jaehyun and Iida, Takeshi},
  journal = {The Journal of Race, Ethnicity, and Politics},
  year    = {2025},
  pages   = {1--26},
  url     = {https://www.cambridge.org/core/journals/journal-of-race-ethnicity-and-politics/article/transnational-political-participation-of-undocumented-mexican-immigrants-in-the-us-respondentdriven-sampling-with-the-hardtoreach-population/25F3F10D14F9C27221F3869A2B7863E1}
}

@article{Assaf2025,
  title   = {Illicit Substance Use and Treatment Access Among Adults Experiencing Homelessness},
  author  = {Assaf, Ryan D. and Morris, Meghan D. and Straus, Elana R. and Martinez, Priest and Philbin, Morgan M. and Kushel, Margot},
  journal = {JAMA},
  year    = {2025},
  volume  = {333},
  number  = {14},
  pages   = {1222--1231},
  doi     = {10.1001/jama.2024.27922},
  url     = {https://jamanetwork.com/journals/jama/fullarticle/2830616}
}

@article{Leal2025,
  title = {Determinants of self-rated health among {V}enezuelan migrant women in {B}razil: {A} cross-sectional study},
  author = {Leal, Maria Carmo and Carvalho, Thaiza Dutra Gomes and Santos, Yamm\^{e} Ramos Portella and Queiroz, Rita Suely Bacuri and Fonseca, Paula Andrea Morelli and Silva, Ant\^{o}nio Augusto Moura and Szwarcwald, Celia Landmann and Riggirozzi, P\'{i}a},
  journal = {The Lancet Regional Health - Americas},
  volume = {45},
  pages = {101077},
  year = {2025},
  doi = {10.1016/j.lana.2025.101077},
  url = {https://doi.org/10.1016/j.lana.2025.101077}
}

@article{Rudolph2024,
  title={Evaluation of respondent-driven sampling in seven studies of people who use drugs from rural populations: {F}indings from the {Rural Opioid Initiative}},
  author={Rudolph, Abby E. and Nance, Robin M. and Bobashev, Georgiy and Brook, Daniel and Akhtar, Wajiha and Cook, Ryan and Cooper, Hannah L. and Friedmann, Peter D. and Frost, Simon D. W. and Go, Vivian F. and Jenkins, Wiley D. and Korthuis, Philip T. and Miller, William C. and Pho, Mai T. and Ruderman, Stephanie A. and Seal, David W. and Stopka, Thomas J. and Westergaard, Ryan P. and Young, April M. and Zule, William A. and Tsui, Judith I.},
  journal={BMC Medical Research Methodology},
  volume={24},
  pages={94},
  number={1},
  year={2024},
  publisher={Springer Nature},
  doi={10.1186/s12874-024-02206-5}
}

@article{Yamanis2013,
  title={An Empirical Analysis of the Impact of Recruitment Patterns on {RDS} Estimates among a Socially Ordered Population of Female Sex Workers in {C}hina},
  author={Yamanis, Thespina J. and Merli, M. Giovanna and Neely, William Whipple and Tian, Felicia Feng and Moody, James and Tu, Xiaowen and Gao, Ersheng},
  journal={Sociological Methods \& Research},
  year={2013},
  volume={42},
  number={3},
  pages={392--425},
  doi={10.1177/0049124113494576},
  pmid={24288418},
  pmcid={PMC3840895}
}

@article{Li2018,
  title = {Overlooked Threats to Respondent Driven Sampling Estimators: {P}eer Recruitment Reality, Degree Measures, and Random Selection Assumption},
  author = {Li, Jianghong and Valente, Thomas W. and Shin, Hee-Sung and Weeks, Margaret and Zelenev, Alexei and Moothi, Gayatri and Mosher, Heather and Heimer, Robert and Robles, Eduardo and Palmer, Greg and Obidoa, Chinekwu},
  journal = {AIDS and Behavior},
  volume = {22},
  number = {7},
  pages = {2340--2359},
  year = {2018},
  doi = {10.1007/s10461-017-1827-1}
}

@article{Heckathorn2011,
  title = {Snowball versus Respondent-Driven Sampling},
  author = {Heckathorn, Douglas D.},
  journal = {Sociological Methodology},
  volume = {41},
  number = {1},
  pages = {355--366},
  year = {2011},
  doi = {10.1111/j.1467-9531.2011.01244.x}
}

@article{Avery2023,
  title   = {Evaluation of Respondent-Driven Sampling Prevalence Estimators Using Real-World Reported Network Degree},
  author  = {Avery, Lisa and Rotondi, Michael},
  journal = {Sociological Methodology},
  volume  = {53},
  number  = {2},
  pages   = {269--287},
  year    = {2023},
  doi     = {10.1177/00811750231163832}
}

@article{Heckathorn2007,
  title   = {Extensions of Respondent-Driven Sampling: {A}nalyzing Continuous Variables and Controlling for Differential Recruitment},
  author  = {Heckathorn, Douglas D.},
  journal = {Sociological Methodology},
  volume  = {37},
  number  = {1},
  pages   = {151--208},
  year    = {2007},
  doi     = {10.1111/j.1467-9531.2007.00188.x}
}

@article{Jayaweera2025,
  author  = {Jayaweera, Ruvani and Odhoch, Lilian and Nabunje, Juliet and Oduor, Clement and Zuniga, Carmela and Powell, Bill and Barasa, Winnie and Aber, Faith and Nyalwal, Bol and Wado, Yadeta Dessalegn and Ouedraogo, Ramatou and Kakesa, Joyce and Fetters, Tamara},
  title   = {Incidence and safety of abortion in two humanitarian settings in {U}ganda and {K}enya: {A} respondent-driven sampling study},
  journal = {eClinicalMedicine},
  volume  = {83},
  pages   = {103200},
  year    = {2025},
  doi     = {10.1016/j.eclinm.2025.103200},
  publisher = {Elsevier}
}

@article{McCreesh2012,
  author  = {McCreesh, Nicky and Frost, Simon D. W. and Seeley, Janet and Katongole, Joseph and Tarsh, Michael N. and Ndunguse, Richard and Jichi, Fred and Lunel, Nelson L. and Maher, Dermot},
  title   = {Evaluation of Respondent-Driven Sampling},
  journal = {Epidemiology},
  year    = {2012},
  volume  = {23},
  number  = {1},
  pages   = {138--147},
  doi     = {10.1097/EDE.0b013e31823ac17c}
}

@article{Heckathorn2002,
  author  = {Heckathorn, Douglas D.},
  title   = {Respondent-Driven Sampling {II}: {D}eriving Valid Population Estimates from Chain-Referral Samples of Hidden Populations},
  journal = {Social Problems},
  volume  = {49},
  number  = {1},
  pages   = {11--34},
  year    = {2002},
  doi     = {10.1525/sp.2002.49.1.11}
}

@article{Wirtz2021,
  author  = {Wirtz, Andrea L. and Iyer, Jessica and Brooks, Durryle and Hailey-Fair, Kimberly and Galai, Noya and Beyrer, Chris and Celentano, David and Arrington-Sanders, Renata},
  title   = {An Evaluation of Assumptions Underlying Respondent-Driven Sampling and the Social Contexts of Sexual and Gender Minority Youth Participating in {HIV} Clinical Trials in the {U}nited {S}tates},
  journal = {Journal of the International AIDS Society},
  year    = {2021},
  volume  = {24},
  number  = {5},
  pages   = {e25694},
  doi     = {10.1002/jia2.25694}
}

@article{Tourangeau2014,
  author  = {Tourangeau, Roger and Edwards, Brad and Johnson, Timothy},
  title   = {Understanding Respondent-Driven Sampling from a Total Survey Error Perspective},
  journal = {Survey Practice},
  volume  = {7},
  number  = {2},
    pages   = {1--6},
  year    = {2014}
}

@article{Fellows2022,
  author  = {Fellows, Ian E.},
  title   = {On the Robustness of Respondent-Driven Sampling Estimators to Measurement Error},
  journal = {Journal of Survey Statistics and Methodology},
  volume  = {10},
  number  = {2},
  pages   = {377--396},
  year    = {2022},
  doi     = {10.1093/jssam/smab056}
}

@article{Fellows2019,
  author  = {Fellows, Ian E.},
  title   = {Respondent-Driven Sampling and the Homophily Configuration Graph},
  journal = {Statistics in Medicine},
  volume  = {38},
  number  = {1},
  pages   = {131--150},
  year    = {2019},
  doi     = {10.1002/sim.7973}
}

@article{Barash2016,
  author  = {Barash, Vladimir and Cameron, Christopher and Heckathorn, Douglas},
  title   = {Respondent-Driven Sampling: {T}esting Assumptions},
  journal = {Journal of Official Statistics},
  volume  = {32},
  number  = {1},
  pages   = {29--73},
  year    = {2016},
  doi     = {10.1515/jos-2016-0002}
}

@article{Wang2024,
  author  = {Wang, Peng and Wei, Chongyi and McFarland, Willi and Raymond, Henry F.},
  title   = {The Development and the Assessment of Sampling Methods for Hard-to-Reach Populations in {HIV} Surveillance},
  journal = {Journal of Urban Health},
  year    = {2024},
  volume  = {101},
  number  = {4},
  pages   = {856--866},
  doi     = {10.1007/s11524-024-00880-w},
  pmid    = {38787451},
  pmcid   = {PMC11329483},
  url     = {https://pmc.ncbi.nlm.nih.gov/articles/PMC11329483/}
}

@article{PAP2024,
  author  = {Fonseca de Barros, Bruna and Fynn, Inés  and Nocetto, Lihuen and Beaudry, Isabelle and  Luna, Juan Pablo and Piñeiro, Rafael and Rosenblatt Rodríguez, Fernando},
  title   = {How parties take advantage of immigrant waves. {P}olitical incorporation in {C}hile},
  journal = {Open Science Framework},
  year    = {2024},
  url     = {osf.io/ab2h5}
}

@article{Roch2018,
  title={Generalized least squares can overcome the critical threshold in respondent-driven sampling},
  author={Roch, Sebastien and Rohe, Karl},
  journal={Proceedings of the National Academy of Sciences},
  volume={115},
  number={41},
  pages={10299--10304},
  year={2018},
  doi={10.1073/pnas.1706699115}
}

\appendix

\section{Apendix}\label{appendix}

\newcolumntype{L}[1]{>{\raggedright\arraybackslash}p{#1}}

\begin{table}[ht!]
\centering
\caption{Standard deviation (SD) across estimators and scenarios.}
\begin{tabular}{lcccccc}
\toprule 
\multirow{2}{*}{Scenario = $(\tau,\phi^{\text{MDR}})$} & \multicolumn{6}{c}{Estimator} \\
\cmidrule{2-7}
& $\hat{\mu}_{VH}^{II}$ & $\hat{\mu}_{DR}^{II}$ & $\hat{\mu}_{MDR}^{II}$ & $\hat{\mu}_{Lu}^{ego}$ & $\hat{\mu}_{DR}^{ego}$ & $\hat{\mu}_{MDR}^{ego}$ \\ 
\midrule
1 : (None, None) & 0.0286 & 0.0188 & 0.0190 & 0.0158 &  0.0157 & 0.0157  \\

2 : (None, Moderate) & 0.0306 &  0.0187 & 0.0203 & 0.0153 & 0.0149 & 0.0164   \\

3 : (None, High) & 0.0294 & 0.0177 & 0.0207 & 0.0144  & 0.0142 & 0.0161  \\

4 : (Moderate, None) & 0.0346 &  0.0277 & 0.0286 & 0.0249 & 0.0248 & 0.0250 \\

5 : (Moderate, Moderate) & 0.0470 & 0.0354 & 0.0308 & 0.0319 & 0.0305 & 0.0252 \\

6 : (Moderate, High) & 0.0521 &  0.0398 & 0.0358 & 0.0343 & 0.0328 & 0.0285 \\

7 : (High, None) & 0.0342  & 0.0284 & 0.0309 & 0.0255 & 0.0256 & 0.0273  \\

8 : (High, Moderate) & 0.0446 & 0.0382 & 0.0343 &0.0332 & 0.0328 & 0.0282 \\

9 : (High, High) & 0.0475 & 0.0422 & 0.0394 & 0.0353 & 0.0353 & 0.0309\\
\bottomrule
\end{tabular}
\label{tab:SD}
\end{table}

\begin{table}[ht!]
\centering
\caption{Error estimation or bias across estimators and scenarios.}
\begin{tabular}{lcccccc}
\toprule 
\multirow{2}{*}{Scenario = $(\tau,\phi^{\text{MDR}})$} & \multicolumn{6}{c}{Estimator} \\
\cmidrule{2-7}
& $\hat{\mu}_{VH}^{II}$ & $\hat{\mu}_{DR}^{II}$ & $\hat{\mu}_{MDR}^{II}$ & $\hat{\mu}_{Lu}^{ego}$ & $\hat{\mu}_{DR}^{ego}$ & $\hat{\mu}_{MDR}^{ego}$ \\ 
\midrule
1 : (None, None) & 0.0013 & 0.0010 &  0.0012 &  0.0003 &  0.0003 &  0.0002  \\

2 : (None, Moderate) & 0.0779  & 0.0052 &  0.0054 & 0.0070 & 0.0012  & 0.0014\\

3 : (None, High) & 0.1035 & 0.0083 & 0.0135 & 0.0088 & 0.0012 &  0.0028  \\

4 : (Moderate, None) & 0.0008 &  0.0006 & -0.0001 & 0.0006  & 0.0005 & -0.0005 \\

5 : (Moderate, Moderate)&  0.1051 &  0.0438 &  0.0125 & 0.0436 &   0.0367  & 0.0077 \\

6 : (Moderate, High) & 0.1384 &  0.0639 &  0.0389 & 0.0581  & 0.0492  & 0.0277\\

7 : (High, None) & 0.0014 &  0.0001 &-0.0003 &  0.0006  & 0.0005 & -0.0003 \\

8 : (High, Moderate) & 0.1111 & 0.0511 &  0.0139 & 0.0525  & 0.0450 &  0.0097 \\

9 : (High, High) & 0.1472 & 0.0737 & 0.0457 & 0.0700 &  0.0599  & 0.0361\\
\bottomrule
\end{tabular}
\label{tab:SESGO}
\end{table}

\begin{result}\label{res:LuyVH}
For $\hat{\mu}_{MDR}^{II}$ and $\hat{\mu}_{MDR}^{ego}$ described in Section \ref{sec:MDRprevalence}, we have that:
\begin{equation}
\hat{\mu}_{MDR}^{ego}
=
\frac{\hat{\mu}_{MDR}^{II}}{
\hat{\mu}_{MDR}^{II} + (1 - \hat{\mu}_{MDR}^{II})\, c
},
\quad \text{where} \quad
c =
\frac{\sum_{i=1}^{N} S_i z_i d_{i0} / \hat{\pi}^{MDR}_i}{
\sum_{i=1}^{N} S_i (1 - z_i) d_{i1} / \hat{\pi}^{MDR}_i
}.
\end{equation}
\end{result}

\textbf{Proof:}
We have:
\begin{equation}
\hat{\mu}_{MDR}^{ego}
=
\frac{
\widehat{C}_{0,1}^{ego,MDR}\widehat{D}_{0}^{ego,MDR}
}{
\widehat{C}_{0,1}^{ego,MDR}\widehat{D}_{0}^{ego,MDR}
+
\widehat{C}_{1,0}^{ego,MDR}\widehat{D}_{1}^{ego,MDR}
}.
\label{eq:mu_ego_mdr}
\end{equation}
Substituting \eqref{eq:ego_mdr1}--\eqref{eq:ego_mdr2} and simplifying:
\[
\hat{\mu}_{MDR}^{ego}
=
\frac{
\frac{\sum_i S_i (1-z_i)d_{i1}/\hat{\pi}_i}{\sum_i S_i (1-z_i)/\hat{\pi}_i}
}{
\frac{\sum_i S_i (1-z_i)d_{i1}/\hat{\pi}_i}{\sum_i S_i (1-z_i)/\hat{\pi}_i}
+
\frac{\sum_i S_i z_i d_{i0}/\hat{\pi}_i}{\sum_i S_i z_i/\hat{\pi}_i}
}
=
\frac{1}{1+
\frac{\sum_i S_i z_i d_{i0}/\hat{\pi}_i}{\sum_i S_i z_i/\hat{\pi}_i}
\frac{\sum_i S_i (1-z_i)/\hat{\pi}_i}{\sum_i S_i (1-z_i)d_{i1}/\hat{\pi}_i}
}
\]
\[
=
\frac{
\frac{\sum_i S_i z_i/\hat{\pi}_i}{\sum_i S_i/\hat{\pi}_i}
}{
\frac{\sum_i S_i z_i/\hat{\pi}_i}{\sum_i S_i/\hat{\pi}_i}
+
\frac{\sum_i S_i (1-z_i)/\hat{\pi}_i}{\sum_i S_i/\hat{\pi}_i}
\frac{\sum_i S_i z_i d_{i0}/\hat{\pi}_i}{\sum_i S_i (1-z_i)d_{i1}/\hat{\pi}_i}
}
=
\frac{\hat{\mu}_{MDR}^{II}}{\hat{\mu}_{MDR}^{II}+(1-\hat{\mu}_{MDR}^{II})c},
\quad
c=\frac{\sum_i S_i z_i d_{i0}/\hat{\pi}_i}{\sum_i S_i (1-z_i)d_{i1}/\hat{\pi}_i}. \quad \square
\]

\end{document}